\pdfoutput=1
\newcommand*{\ATLASLATEXPATH}{./}
\documentclass[UKenglish,texlive=2011,cernpreprint]{\ATLASLATEXPATH atlasdoc}
\AtlasJournalRef{Physics Letters B 761 (2016) 350 - 371}
\AtlasDOI{10.1016/j.physletb.2016.08.042}
\usepackage[biblatex=false]{\ATLASLATEXPATH atlaspackage}
\usepackage{\ATLASLATEXPATH atlasphysics}
\AtlasTitle{Measurement of the top quark mass in the \ttbarll\ channel from 
            $\sqrt{s}=8$~\TeV\ ATLAS data}

\author{The ATLAS Collaboration}

 \date{\today}

 \PreprintIdNumber{CERN-EP-2016-114}

\AtlasAbstract{%
 The top quark mass is measured in the \ttbarll\ channel ($\mbox{lepton}=e,\mu$)
 using ATLAS data recorded in the year 2012 at the LHC.
 The data were taken at a proton--proton centre-of-mass energy of
 $\sqrts=8$~\TeV\ and correspond to an integrated luminosity of
 about \atlumo~\ifb.
 Exploiting the template method, and using the distribution of invariant masses
 of lepton--\bjet\ pairs, the top quark mass is measured to be
 $\mt=\XZ{\EightValue}{\EightStat}{\EightSyst}$\,\GeV, with a total uncertainty
 of \EightTot~\GeV.
 Finally, a combination with previous ATLAS \mt\ measurements from
 $\sqrts=7$~\TeV\ data in the \ttbarll\ and \ttbarlj\ channels results in
 $\mt= \XZ{\CombValue}{\CombStat}{\CombSyst}$\,\GeV, with a total uncertainty
 of \CombTot~\GeV.
}

\hypersetup{pdftitle={ATLAS document},pdfauthor={The ATLAS Collaboration}}
\usepackage{rotating}
\usepackage{cite}
\newcommand{\atlumo                    }{\ensuremath{20.2}\xspace}
\newcommand{\atlumounc                 }{\ensuremath{1.9\%}\xspace}
\newcommand{\xsecunc                   }{\ensuremath{5.4\%}\xspace}
\newcommand{\stopxsecunc               }{\ensuremath{6.0\%}\xspace}
\newcommand{\JESunc                    }{\ensuremath{4.1\%}\xspace}
\newcommand{\bJESunc                   }{\ensuremath{2.2\%}\xspace}
\newcommand{\btagunc                   }{\ensuremath{2.8\%}\xspace}
\newcommand{\qcdunc                    }{\ensuremath{100\%}\xspace}
\newcommand{\eff                       }{\ensuremath{78.4}\xspace}
\newcommand{\effunc                    }{\ensuremath{0.2}\xspace}
\newcommand{\fracmatch                 }{\ensuremath{51.6}\xspace}
\newcommand{\fracmatchunc              }{\ensuremath{0.1}\xspace}
\newcommand{\fracunmatch               }{\ensuremath{34.2}\xspace}
\newcommand{\fracunmatchunc            }{\ensuremath{0.1}\xspace}
\newcommand{\fracwrong                 }{\ensuremath{14.2}\xspace}
\newcommand{\fracwrongunc              }{\ensuremath{0.1}\xspace}
\newcommand{\Cutmatcheff               }{\ensuremath{95.3}\xspace}
\newcommand{\Cutmatcheffunc            }{\ensuremath{0.4}\xspace}
\newcommand{\Cutsignalpurity           }{\ensuremath{69.8}\xspace}
\newcommand{\Cutsignalpurityunc        }{\ensuremath{0.3}\xspace}
\newcommand{\fracunmatchopti           }{\ensuremath{26.7}\xspace}
\newcommand{\fracunmatchoptiunc        }{\ensuremath{0.1}\xspace}
\newcommand{\fracwrongopti             }{\ensuremath{3.4}\xspace}
\newcommand{\fracwrongoptiunc          }{\ensuremath{0.0}\xspace}
\newcommand{\OptiFraction              }{\ensuremath{26}\xspace}
\newcommand{\bkgfrEightTeV             }{\ensuremath{0.01}\xspace}
\newcommand{\bkgfrEightTeVunc          }{\ensuremath{0.00}\xspace}
\newcommand{\ExpstatuncEight           }{\ensuremath{0.41}\xspace}
\newcommand{\ExpstatuncEightunc        }{\ensuremath{0.03}\xspace}
\newcommand{\EightValue                }{\ensuremath{172.99}\xspace}
\newcommand{\EightStat                 }{\ensuremath{0.41}\xspace}
\newcommand{\EightSyst                 }{\ensuremath{0.74}\xspace}
\newcommand{\EightTot                  }{\ensuremath{0.84}\xspace}
\newcommand{\EightImprovementSeven     }{\ensuremath{40\%}\xspace}
\newcommand{\difffit                   }{\ensuremath{0.2}\xspace}
\newcommand{\CombValue                 }{\ensuremath{172.84}\xspace}
\newcommand{\CombStat                  }{\ensuremath{0.34}\xspace}
\newcommand{\CombSyst                  }{\ensuremath{0.61}\xspace}
\newcommand{\CombTot                   }{\ensuremath{0.70}\xspace}
\newcommand{\CombImprovement           }{\ensuremath{17\%}\xspace}
\newcommand{\CombChiSqProb             }{\ensuremath{73\%}\xspace}
\newcommand{\CombWeightLLseven         }{\ensuremath{0.30}\xspace}
\newcommand{\CombWeightDLseven         }{\ensuremath{0.07}\xspace}
\newcommand{\CombWeightDLeight         }{\ensuremath{0.63}\xspace}
\newcommand{\CompatibDil               }{\ensuremath{0.66}\xspace}
\newcommand{\CompatibMixed             }{\ensuremath{0.43}\xspace}
\newcommand{\CompatibSeven             }{\ensuremath{0.75}\xspace}
\newcommand{\CorrDil                   }{\ensuremath{0.51}\xspace}
\newcommand{\CorrEightDilSevenLj       }{\ensuremath{0.00}\xspace}

\newcommand{\CombValueStab             }{\ensuremath{0.03}\xspace}
\newcommand{\CombValueDL               }{\ensuremath{173.04}\xspace}
\newcommand{\CombStatDL                }{\ensuremath{0.38}\xspace}
\newcommand{\CombSystDL                }{\ensuremath{0.74}\xspace}
\newcommand{\CombTotDL                 }{\ensuremath{0.84}\xspace}
\newcommand{\CombChiSqProbDL           }{\ensuremath{51\%}\xspace}
\newcommand{\CombWeightDLeightDL       }{\ensuremath{0.94}\xspace}
\newcommand{\CombValueStabDL           }{\ensuremath{0.03}\xspace}
\newcommand{\CombTotStabDL             }{\ensuremath{0.04}\xspace}
\newcommand{\threed       }{\ensuremath{\mbox{three-dimensional}}\xspace}
\newcommand{\SelPurity    }{Selection purity}
\newcommand{\selPurity    }{selection purity}
\newcommand{\XZ           }[3]{\ensuremath{#1\,\pm #2~\mathrm{(stat)}\,\pm #3~\mathrm{(syst)}}\xspace}
\newcommand{\XZst         }[2]{\ensuremath{#1\,\pm #2~\mathrm{(stat)}}\xspace}
\newcommand{\XZtot        }[2]{\ensuremath{#1\,\pm #2}\xspace}
\newcommand{\ttbarll      }{\ensuremath{\ttbar\to\mbox{dilepton}}\xspace}
\newcommand{\ttbarlj      }{\ensuremath{\ttbar\to\mbox{lepton+jets}}\xspace}
\newcommand{\pp           }{\ensuremath{\ensuremath{pp}}\xspace}
\newcommand{\ljets        }{\ensuremath{\ell\mbox{+jets}}\xspace}
\newcommand{\dil          }{\ensuremath{\mbox{dilepton}}\xspace}
\newcommand{\emu          }{\ensuremath{e\mu}\xspace}

\newcommand{\Zj           }{$Z$+jets\xspace}
\newcommand{\mt           }{\ensuremath{m_{\mathrm{top}}}\xspace}
\newcommand{\mtseven      }{\ensuremath{m_{\mathrm{top}}^{\mathrm{7\!\TeV}}}\xspace}
\newcommand{\mtdl         }{\ensuremath{m_{\mathrm{top}}^{\mathrm{dil}}}\xspace}
\newcommand{\dmtdl        }{\ensuremath{\Delta m_{\mathrm{top}}^{\mathrm{dil}}}\xspace}
\newcommand{\mtlj         }{\ensuremath{m_{\mathrm{top}}^{\mathrm{\ljets}}}\xspace}
\newcommand{\mtall        }{\ensuremath{m_{\mathrm{top}}^{\mathrm{all}}}\xspace}
\newcommand{\mlb          }{\ensuremath{m_{\ell b}}\xspace}
\newcommand{\mlbr         }{\ensuremath{m_{\ell b}^{\mathrm{reco}}}\xspace}
\newcommand{\ptlb         }{\ensuremath{p_{\mathrm{T},\ell b}}\xspace}
\newcommand{\dR           }{\ensuremath{\Delta R}\xspace}
\newcommand{\dRlb         }{\ensuremath{\Delta R_{\ell\mathrm{b}}}\xspace}
\newcommand{\hdamp        }{\ensuremath{h_{\mathrm{damp}}}\xspace}
\newcommand{\pb             }{\ensuremath{\mathrm{pb}\xspace}}
\newcommand{\invfb          }{\ensuremath{\mathrm{fb}^{-1}\xspace}}
\newcommand{\cmes           }{centre-of-mass energies\xspace}
\newcommand{\etaclus        }{\ensuremath{\eta_\mathrm{cluster}}\xspace}
\newcommand{\absetaclus     }{\ensuremath{\vert\etaclus\vert}\xspace}
\newcommand{\nvtx         }{\ensuremath{n_\mathrm{vtx}}\xspace}
\newcommand{\meanmu       }{\ensuremath{\langle\mu\rangle}\xspace}
\newcommand{\sqrts        }{\ensuremath{{\sqrt{s}}}\xspace}
\newcommand{\LambdaQCD    }{\ensuremath{\Lambda_\mathrm{QCD}}\xspace}
\newcommand{\btag         }{\ensuremath{b\mbox{-tagging}}\xspace}

\newcommand{\btagged      }{\ensuremath{b\mbox{-tagged}}\xspace}
\newcommand{\bjet         }{\ensuremath{\mathit{b}\mbox{-jet}}\xspace}
\newcommand{\antikt       }{\ensuremath{\mbox{anti-}k_t}\xspace}

\newcommand{\Zbos         }{\ensuremath{Z}\xspace}
\newcommand{\Wbos         }{\ensuremath{W^{\pm}}\xspace}
\newcommand{\Alpgen       }{{\sc Alpgen}\xspace}
\renewcommand{\Herwig       }{{\sc Herwig}\xspace}
\newcommand{\Jimmy        }{{\sc Jimmy}\xspace}
\newcommand{\Mcatnlo      }{{\sc MC@NLO}\xspace}
\newcommand{\Pythia       }{{\sc Pythia}\xspace}
\newcommand{\PowhegHerwig }{{\sc Powheg+Herwig}\xspace}
\newcommand{\PowhegBox    }{{\sc Powheg-Box}\xspace}
\newcommand{\Pythiasix    }{{\sc Pythia6}\xspace}

\newcommand{\Powheg       }{{\sc Powheg}\xspace}
\newcommand{\Geantfour    }{{\sc Geant4}\xspace}
\begin{document}
\maketitle
\section{Introduction}
\label{sec:intro}
 The mass of the top quark~(\mt) is an important parameter of the Standard
 Model~(SM) of particle physics.
 Precise measurements of \mt\ provide crucial information for global fits of
 electroweak parameters~\cite{LEPewkfits, Baak2014,PDG2014} which help assess
 the internal consistency of the SM and to probe its extensions. In addition,
 the value of \mt\ affects the stability of the SM Higgs potential, which has
 cosmological implications~\cite{Degrassi2012, HiggsStab1, HiggsStab2}.
 Many measurements of \mt\ have been performed by the Tevatron and LHC
 collaborations.
 Combining a selection of those, the first Tevatron+LHC \mt\ result is
 $\mt=\XZ{173.34}{0.27}{0.71}$\,\GeV, with a total uncertainty of
 $0.76$~\GeV~\cite{ATLAS:2014wva}.
 Meanwhile, a number of new results have become
 available~\cite{ATL-2015-002,CDF-1501,CDF-1401,CMS-1401,D00-1501,D00-1401},
 some of which are more precise than the above combination.
 The latest ATLAS results in the \ttbarlj\ and \ttbarll\ decay channels, both
 with electrons ($e$) and muons ($\mu$) in the final state~\cite{Aad:2015nba},
 are $\mt=\XZ{172.33}{0.75}{1.02}$\,\GeV\ and
 $\mt=\XZ{173.79}{0.54}{1.30}$\,\GeV, respectively.

 This Letter presents a new measurement of \mt\ obtained in the \ttbarll\ decay
 channel using 2012 data taken at a proton--proton (\pp) centre-of-mass energy
 of $\sqrts=8$~\TeV, with an integrated luminosity of about $\atlumo$~\invfb.
 The analysis exploits the decay
 $\ttbar\to\Wplus\Wminus\bbbar\to\ellell\nnbar\bbbar$, which is realised when
 both \Wboson\ bosons decay into a charged lepton and its corresponding
 neutrino. In the analysis, the \ttbar\ decay channels $ee, e\mu$ and $\mu\mu$
 (including $\tau\to e,\mu$) are combined and referred to as the \dil\ channel.
 Single-top-quark events with the same lepton final states are included in the
 signal.
 Given the larger data sample compared to \Ref{\cite{Aad:2015nba}}, the event
 selection was optimised to achieve the smallest total uncertainty.
 The measurement is based on the implementation of the template method described
 in \Ref{\cite{Aad:2015nba}}, which is calibrated using signal Monte Carlo~(MC)
 samples. Consequently, the top quark mass measured in this way corresponds to
 the mass definition used in the MC program.

\section{ATLAS detector}
\label{sec:atlas}
 \newcommand{\AtlasCoordFootnote}{%
  ATLAS uses a right-handed coordinate system
  with its origin at the nominal interaction point (IP) in the centre of the
  detector and the $z$-axis along the beam pipe.  The $x$-axis points from the
  IP to the centre of the LHC ring, and the $y$-axis points upwards.
  Cylindrical coordinates $(r,\phi)$ are used in the transverse plane, $\phi$
  being the azimuthal angle around the $z$-axis.  The pseudorapidity is defined
  in terms of the polar angle $\theta$ as $\eta = -\ln \tan(\theta/2)$.  Angular
  distance is measured in units of $\Delta R \equiv \sqrt{(\Delta\eta)^{2} +
    (\Delta\phi)^{2}}$.}
 The ATLAS experiment~\cite{PERF-2007-01} at the LHC is a multi-purpose particle
 detector with a forward-backward symmetric cylindrical geometry and a near
 $4\pi$ coverage in solid angle.\footnote{\AtlasCoordFootnote} It consists of an
 inner tracking detector surrounded by a thin superconducting solenoid providing
 a \SI{2}{\tesla} axial magnetic field, electromagnetic and hadron calorimeters,
 and a muon spectrometer.  The inner tracking detector covers the pseudorapidity
 range $|\eta| < 2.5$.  It consists of silicon pixel, silicon microstrip, and
 transition radiation tracking detectors.  Lead/liquid-argon (LAr) sampling
 calorimeters provide electromagnetic (EM) energy measurements with high
 granularity.  A hadronic (steel/scintillator-tile) calorimeter covers the
 central pseudorapidity range ($|\eta| < 1.7$).  The end-cap and forward regions
 are instrumented with LAr calorimeters for EM and hadronic energy
 measurements up to $|\eta| = 4.9$.  The muon spectrometer surrounds the
 calorimeters and is based on three large air-core toroid superconducting
 magnets with eight coils each.  Its bending power is in the range from
 \num{2.0} to \SI{7.5}{\tesla\metre}.  It includes a system of precision
 tracking chambers and fast detectors for triggering.  A three-level trigger
 system is used to select events.  The first-level trigger is implemented in
 hardware and uses a subset of the detector information to reduce the accepted
 event rate to at most \SI{75}{\kilo\hertz}.  This is followed by two
 software-based trigger levels that together reduce the accepted rate to
 \SI{400}{\hertz} on average depending on the data-taking conditions during
 2012.

\section{Data and MC samples}
\label{sec:mc}
 This analysis is based on \pp\ collision data recorded in 2012 at
 $\sqrts=8$~\TeV.
 The integrated data luminosity amounts to $\atlumo$~\invfb\ with an uncertainty
 of \atlumounc\ determined with the procedures described
 in \Ref{\cite{Aad:2013ucp}}.

 The modelling of \ttbar\ and single-top-quark signal events and of most
 background processes relies on MC simulations.
 For the simulation of signal events the \PowhegBox\
 program~\cite{NAS-0401,FRI-0701,Alioli:2010xd} is used.
 The simulation of the top quark pair~\cite{FRI-0702} and single-top-quark
 production in the $Wt$-channel~\cite{REE-1101} uses matrix elements at
 next-to-leading order~(NLO) in the strong coupling constant \alphas, with the
 NLO~CT10~\cite{LAI-1001} parton distribution function~(PDF) and the parameter
 $\hdamp=\infty$.  The \hdamp\ parameter sets the resummation scale, which
 controls the transition from the matrix element to the parton shower~(PS)
 simulation.
 Given that the event selection described below requires leptonic decay products
 of two \Wboson\ bosons, single-top-quark events in the $s$-channel and
 $t$-channel are found not to contribute to the sample.

 The \Pythia~(v6.425) program~\cite{SJO-0601} with the P2011C~\cite{Skands} set
 of tuned parameters (tune) and the corresponding CTEQ6L1 PDF~\cite{Pumplin2002}
 are employed to provide the parton shower, hadronisation and underlying-event
 modelling.
 The uncertainties due to QCD initial- and final-state radiation (ISR/FSR)
 modelling are estimated with samples generated with the \PowhegBox\ program
 interfaced to the \Pythia\ program for which the parameters of the generation
 are varied to span the ranges compatible with the results of measurements
 of \ttbar\ production in association with
 jets~\cite{ATL-2012-033,ATLASCollaboration2015,ATL-PHYS-PUB-2015-002}.

 For \mt\ hypothesis testing, the \ttbar\ and single-top-quark event samples are
 generated for five values of \mt\ in the range $167.5$ to $177.5$~\GeV\ in
 steps of $2.5$~\GeV.
 For each \mt\ value, the MC samples are normalised according to the best
 available cross-section calculations, which for $\mt=172.5$~\GeV\ are
 $\sigma_{\ttbar}=253^{+13}_{-15}$~\pb~\cite{CAC-1101,PRL-109-132001,JHEP-1212,JHEP-1301,Czakon:2013goa,CZA-1101}
 for \ttbar\ production and $\sigma_{Wt}=22.4\pm1.5$~\pb~\cite{Kidonakis:2010ux}
 for single-top-quark production in the $Wt$-channel.
 The PDF+\alphas-induced uncertainties in these cross-sections are calculated
 using the PDF4LHC prescription~\cite{PDF4LHC} with the MSTW2008 $68\%$ CL~NNLO
 PDF~\cite{MAR-0901,MAR-0902}, CT10~NNLO~PDF\cite{LAI-1001,CT10NNLO} and
 NNPDF2.3~5f~FFN PDF~\cite{Ball:2012cx}, and are added in quadrature with the
 uncertainties due to the choices of the factorisation and renormalisation
 scales.

 The simulation of \Wbos\ or \Zbos\ boson production in association with jets is
 performed with the \Alpgen~(v2.13) program~\cite{MAN-0301} interfaced to
 the \Pythiasix\ program using the CTEQ6L1 PDF and the corresponding AUET2
 tune~\cite{ATL-PHYS-PUB-2011-008}.
 Diboson production processes ($WW$, $WZ$ and $ZZ$) are simulated using
 the \Alpgen\ program interfaced to the \Herwig~(v6.520) program~\cite{COR-0001}
 with the AUET2 tune and to the \Jimmy~(v4.31) program~\cite{SAMPLES-JIMMY}.
 All samples are simulated taking into account the effects of multiple soft \pp\
 interactions (pile-up) registered in the 2012 data. These interactions are
 modelled by overlaying simulated hits from events with exactly one inelastic
 (signal) collision per bunch crossing with hits from minimum-bias events that
 are produced with the \Pythia~(v8.160) program~\cite{SJO-0801} using the A2M
 tune~\cite{ATL-PHYS-PUB-2012-003} and the MSTW2008~LO~PDF.
 For this analysis, the observed values of the pile-up-related
 quantities \meanmu, the mean number of interactions per bunch crossing,
 and \nvtx, the average number of vertices per event, are $\meanmu=20.7$ and
 $\nvtx=9.2$.

 Finally, the samples undergo a simulation of the ATLAS
 detector~\cite{WT-ATLAS-SIMULATION-PAPER} based on \Geantfour~\cite{AGO-0301},
 and are then processed through the same reconstruction software as the data.
 A number of samples used to assess systematic uncertainties are produced with a
 faster version of the simulation which, in addition to the full simulation of
 the tracking, uses smearing functions and interpolates particle behaviour and
 calorimeter response, based on resolution functions measured in full-simulation
 studies, to approximate the results of the full simulation.

\section{Data selection and event reconstruction}
\label{sec:evsel}
 Triggers based on isolated single electrons or muons with energy or momentum
 thresholds of $24$~\GeV\ are used.
 The detector objects resulting from the top quark pair decay are electron and
 muon candidates, jets and missing transverse momentum (\met).
 In the following, the term lepton is used for charged leptons (excluding $\tau$
 leptons) exclusively.

 Electron candidates~\cite{CERN-PH-EP-2014-040} are required to have a
 transverse energy of $\et>25$~\GeV, a pseudorapidity of the corresponding EM
 cluster of $\absetaclus < 2.47$, with the transition region
 $1.37<\absetaclus<1.52$ between the barrel and the end-cap calorimeter
 excluded.
 The muon candidates~\cite{CERN-PH-EP-2014-151} are required to have transverse
 momentum $\pt>25$~\GeV\ and $\vert\eta\vert<2.5$.
 To reduce the contamination by leptons from heavy-flavour decays inside jets or
 from photon conversions, referred to as non-prompt~(NP) leptons, strict
 isolation criteria are applied to the amount of activity in the vicinity of the
 lepton candidate~\cite{CERN-PH-EP-2014-040,CERN-PH-EP-2014-151}.

 Jets are built from topological clusters of calorimeter cells~\cite{Aad:Topo}
 with the \antikt\ jet clustering algorithm~\cite{CAC-0801} using a radius
 parameter of $R=0.4$.
 Jets are reconstructed using the local cluster weighting~(LCW) and global
 sequential calibration~(GSC)
 algorithms~\cite{Aad:1409965,ATLASCollaboration2015b,ATLAS-CONF-2015-037} and
 required to satisfy $\pt>25$~\GeV\ and $\abseta<2.5$.
 Muons reconstructed within a $\dR=0.4$ cone around the axis of a jet with
 $\pt>25$~\GeV\ are not considered as charged-lepton candidates.
 In addition, jets within a $\dR=0.2$ cone around an electron candidate are
 removed and finally electrons within a $\dR=0.4$ cone around any of the
 remaining jets are discarded.
 The identification of jets containing $b$-hadrons, \btag, is used for event
 reconstruction and background suppression.
 In the following, irrespective of their origin, jets tagged by the \btag\
 algorithm are referred to as \btagged\ jets, whereas those not tagged are
 referred to as untagged jets.
 Similarly, whether they are tagged or not, jets originating from bottom quarks
 are referred to as \bjet{s}\ and those from $(u, d, c, s)$-quarks or gluons as
 light jets.
 The working point of the neural-network-based MV1 \btag\
 algorithm~\cite{Aad:2015ydr} corresponds to an average \btag\ efficiency of
 70$\%$ for \bjet{s}\ in simulated \ttbar\ events and rejection factors of 5 for
 jets containing a $c$-hadron and 137 for jets containing only lighter-flavour
 hadrons.
 To match the \btag\ performance in the data, \pt- and $\eta$-dependent scale
 factors~\cite{Aad:2015ydr}, obtained from dijet and \ttbarll\ events, are
 applied to MC jets depending on their true flavour.
 The reconstruction of the \met\ is based on the vector sum of energy deposits
 in the calorimeters, projected onto the transverse plane. Muons are included in
 the \met\ using their reconstructed momentum in the tracking
 detectors~\cite{ATLAS-MET-NEW}.

 The contribution of events wrongly reconstructed as \ttbarll\ events due to the
 presence of objects misidentified as leptons (fake leptons), is estimated from
 data~\cite{ATLAS-CONF-2014-058}.
 The technique employed uses fake-lepton and real-lepton efficiencies that
 depend on $\eta$ and \pt, measured in a background-enhanced control region with
 low \met\ and from events with dilepton masses around the \Zboson\
 peak~\cite{Aad:2010ey}.

 The selection from \Ref{\cite{Aad:2015nba}} is applied as a pre-selection as
 follows:
\begin{enumerate}
 \item Events are required to have a signal from the single-electron or
   single-muon trigger and at least one primary vertex with at least five
   associated tracks.
 \item Exactly two oppositely charged leptons are required, with at least one of
   them matching the reconstructed object that fired the corresponding trigger.
 \item In the same-lepton-flavour channels, $ee$ and $\mu\mu$, $\met>60$~\GeV\
   is required. In addition, the invariant mass of the lepton pair must satisfy
   $m_{\ell\ell} > 15$~\GeV, and must not be compatible with the \Zboson\ mass
   within $10$~\GeV.
 \item In the \emu\ channel the scalar sum of \pt\ of the two selected leptons
   and all jets is required to be larger than $130$~\GeV.
 \item The presence of at least two jets with $\pt>25$~\GeV\ and
   $\vert\eta\vert<2.5$ is required, and at least one of these jets has to
   be \btagged.
\end{enumerate}
 The observed numbers of events in the data after this pre-selection, together
 with the expected numbers of signal and background events corresponding to the
 integrated data luminosity, are given in~\Tab{\ref{tab:selections}}.
 Assuming a top quark mass of $\mt=172.5$~\GeV, the predicted number of events
 is consistent with the one observed in the data within uncertainties.
 For all predictions, the uncertainties are estimated as the sum in quadrature
 of the statistical uncertainty, a \atlumounc\ uncertainty in the integrated
 luminosity, and a number of additional components.
 For the signal, these are a \xsecunc\ uncertainty in the \ttbar\ cross-section,
 or a \stopxsecunc\ uncertainty in the single-top-quark cross-section, as given
 in \Sect{\ref{sec:mc}}.
 Finally, global \JESunc, \bJESunc and \btagunc\ uncertainties are added,
 corresponding to the envelopes of the results from the eigenvector variations
 of the jet energy scale~(JES), the relative $b$-to-light-jet energy
 scale~(bJES) and the \btag\ scale factors, respectively.
 The background uncertainties contain jet-multiplicity-dependent uncertainties
 of about 40$\%$ in the normalisation of the \Zj\ background and a \qcdunc\
 uncertainty in the normalisation of fake-lepton background.
%
\begin{table}[tb!]
\begin{center}
\small
\begin{tabular}{|l|rr|rr|}
\hline
Selection & \multicolumn{2}{c|}{Pre-selection} 
          & \multicolumn{2}{c|}{Final selection}\\  
\hline
Data                         & \multicolumn{2}{c|}{36359} & 
                               \multicolumn{2}{c|}{9426} \\\hline
\ttbar\ signal               & 34300 $\pm$ & 2700 &  9670 $\pm$ &  770 \\
Single-top-quark signal      &  1690 $\pm$ &  110 &   363 $\pm$ &   23 \\
Fake leptons                 &   240 $\pm$ &  240 &    31 $\pm$ &   31 \\
\Zj\                         &   212 $\pm$ &   83 &  20.6 $\pm$ &  8.5 \\
$WW/WZ/ZZ$                   &    57 $\pm$ &   21 &  10.2 $\pm$ &  3.8 \\
Signal+background            & 36600 $\pm$ & 2800 & 10100 $\pm$ &  770 \\ \hline
Expected background fraction & 0.01  $\pm$ & 0.01 & \bkgfrEightTeV $\pm$ & \bkgfrEightTeVunc \\
Data / (Signal+background)   & 0.99  $\pm$ & 0.07 & 0.93 $\pm$ & 0.07 \\ 
\hline
Matching efficiency [\%]    &\eff         $\pm$ &\effunc         &\Cutmatcheff     $\pm$ &\Cutmatcheffunc \\
\SelPurity\ [\%]            &\fracmatch   $\pm$ &\fracmatchunc   &\Cutsignalpurity $\pm$ &\Cutsignalpurityunc \\
Unmatched events [\%]       &\fracunmatch $\pm$ &\fracunmatchunc &\fracunmatchopti $\pm$ &\fracunmatchoptiunc \\
Wrongly matched events [\%] &\fracwrong   $\pm$ &\fracwrongunc   &\fracwrongopti   $\pm$ &\fracwrongoptiunc \\
\hline
\end{tabular}
\end{center}
\caption{The observed numbers of events in data after the pre-selection and
 the final selection.
 In addition, the expected numbers of signal events for $\mt=172.5$~\GeV\ and
 background events corresponding to the integrated data luminosity are given.
 Two significant digits are used for the uncertainties of the predicted numbers
 of events explained in the text.
 The lower rows report the matching performance evaluated for $\mt=172.5$~\GeV,
 using one significant digit for the statistical uncertainties.
\label{tab:selections}}
\end{table}

 The two jets carrying the highest MV1 weight are taken as the two \bjet{s}
 originating from the decays of the two top quarks, and the two leptons are
 taken as the leptons from the leptonic \Wboson\ decays.
 From the two possible assignments of the two pairs, the combination leading to
 the lowest average invariant mass of the two lepton--\bjet\ pairs (\mlb) is
 retained.
 To estimate the performance of this algorithm in MC simulated samples, the
 reconstruction-level objects are matched to the closest generator-level object
 based on a maximum allowed \dR, being $0.1$ for leptons and $0.3$ for jets.
%
%
\begin{figure*}[tbp!]
\centering
\subfloat[]{
  \includegraphics[width=0.48\textwidth]{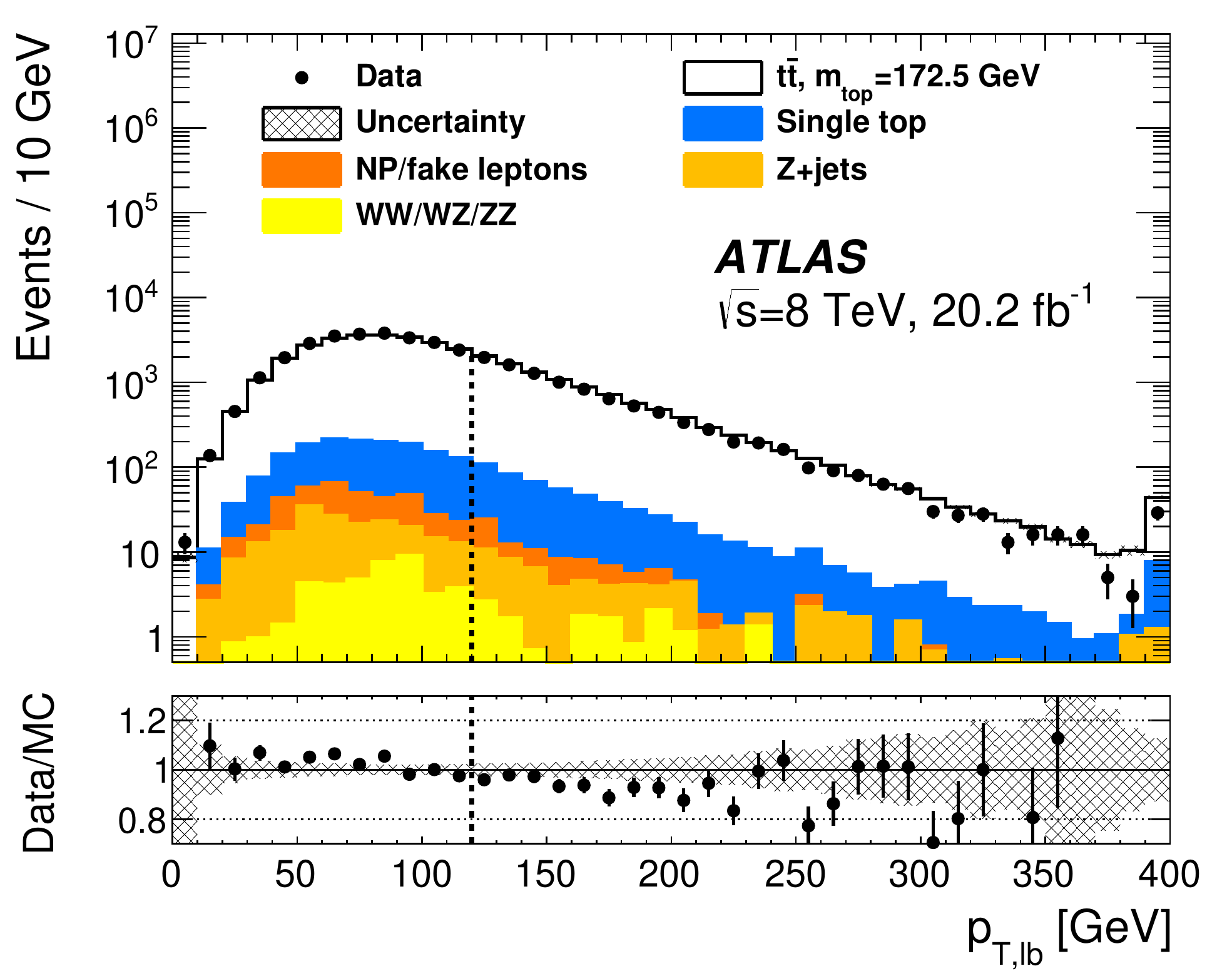}\label{sfig:dataMCptlb}}
\subfloat[]{
  \includegraphics[width=0.48\textwidth]{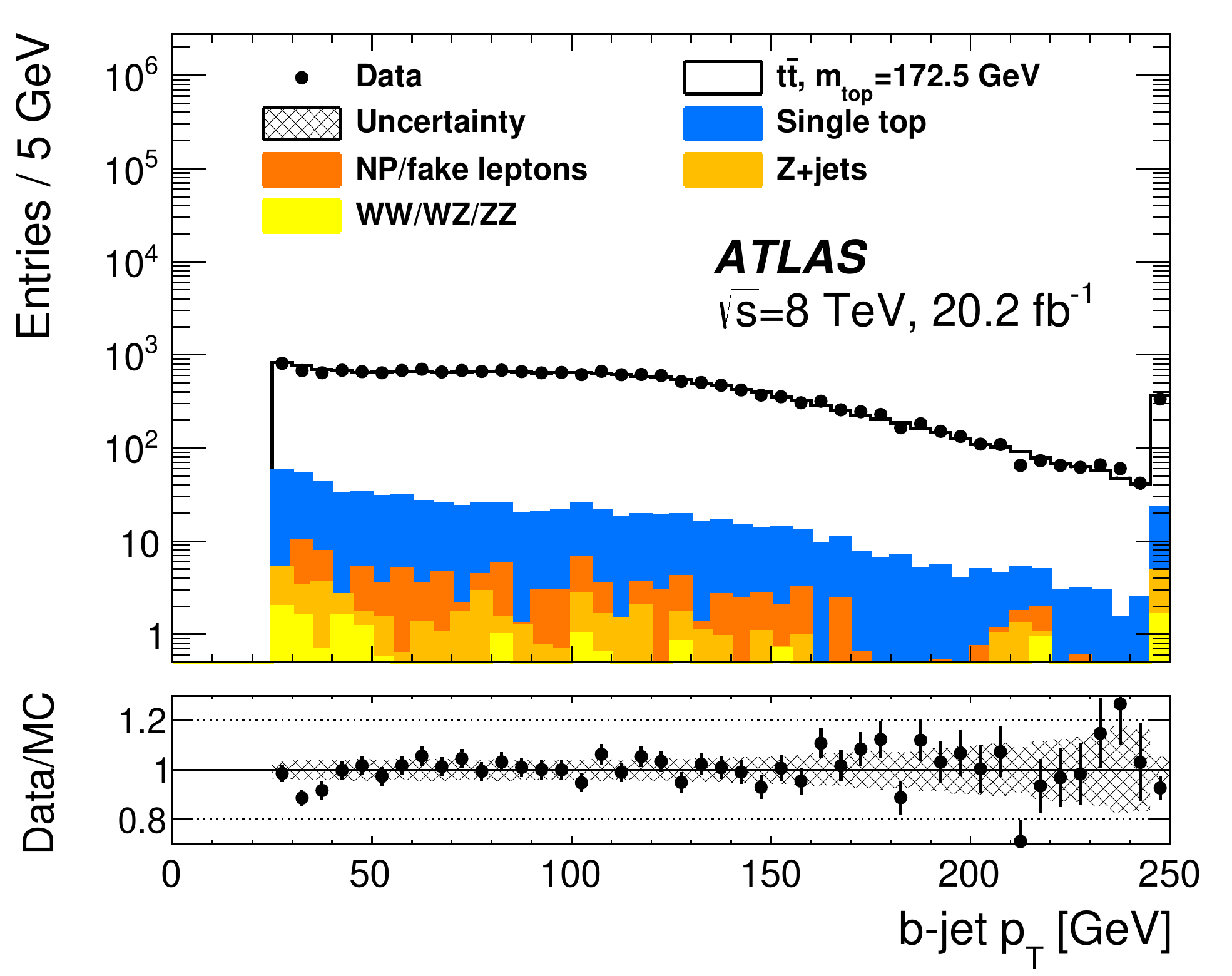}\label{sfig:dataMCptbj}}
\hfill
\subfloat[]{
  \includegraphics[width=0.48\textwidth]{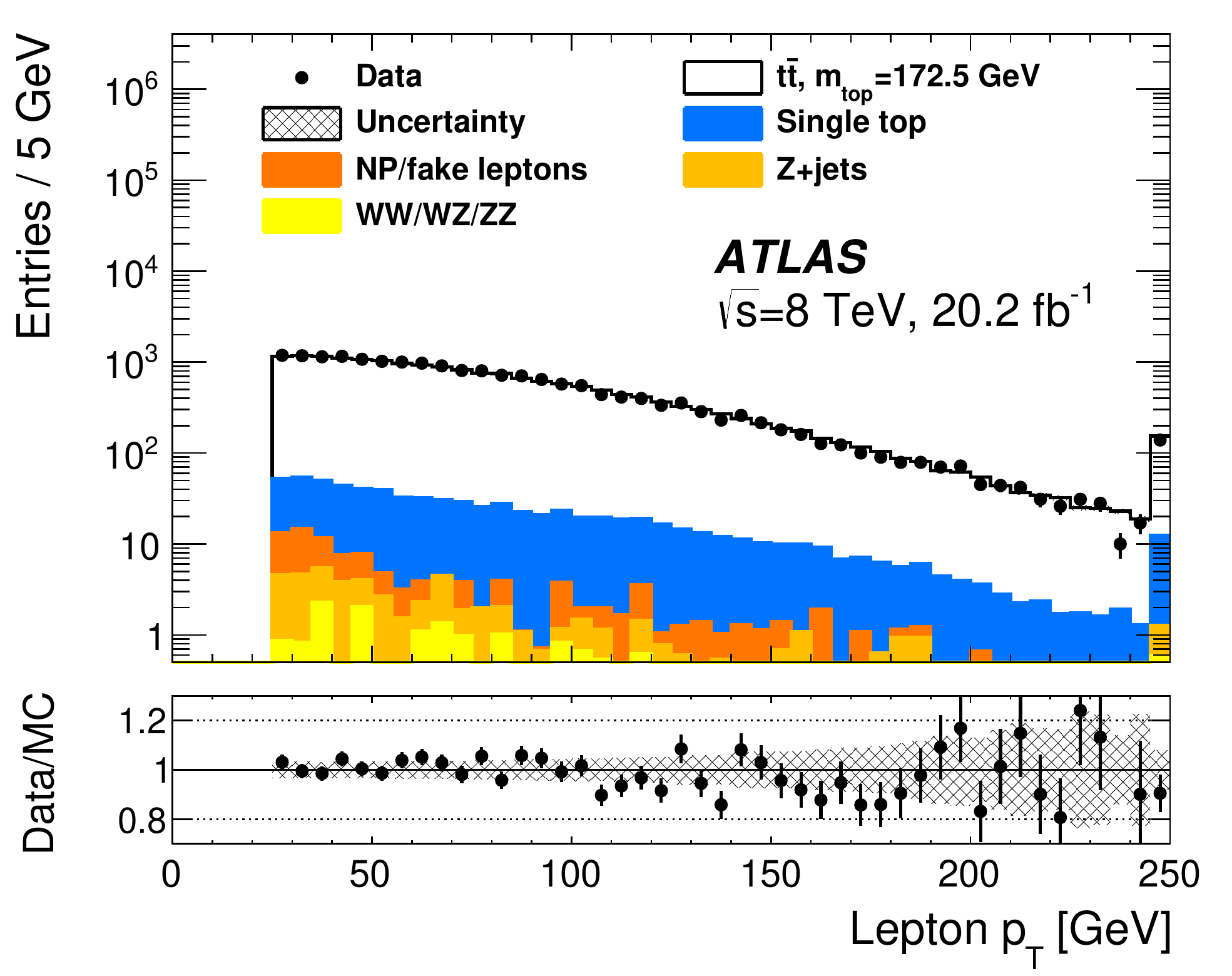}\label{sfig:dataMCptle}}
\subfloat[]{
  \includegraphics[width=0.48\textwidth]{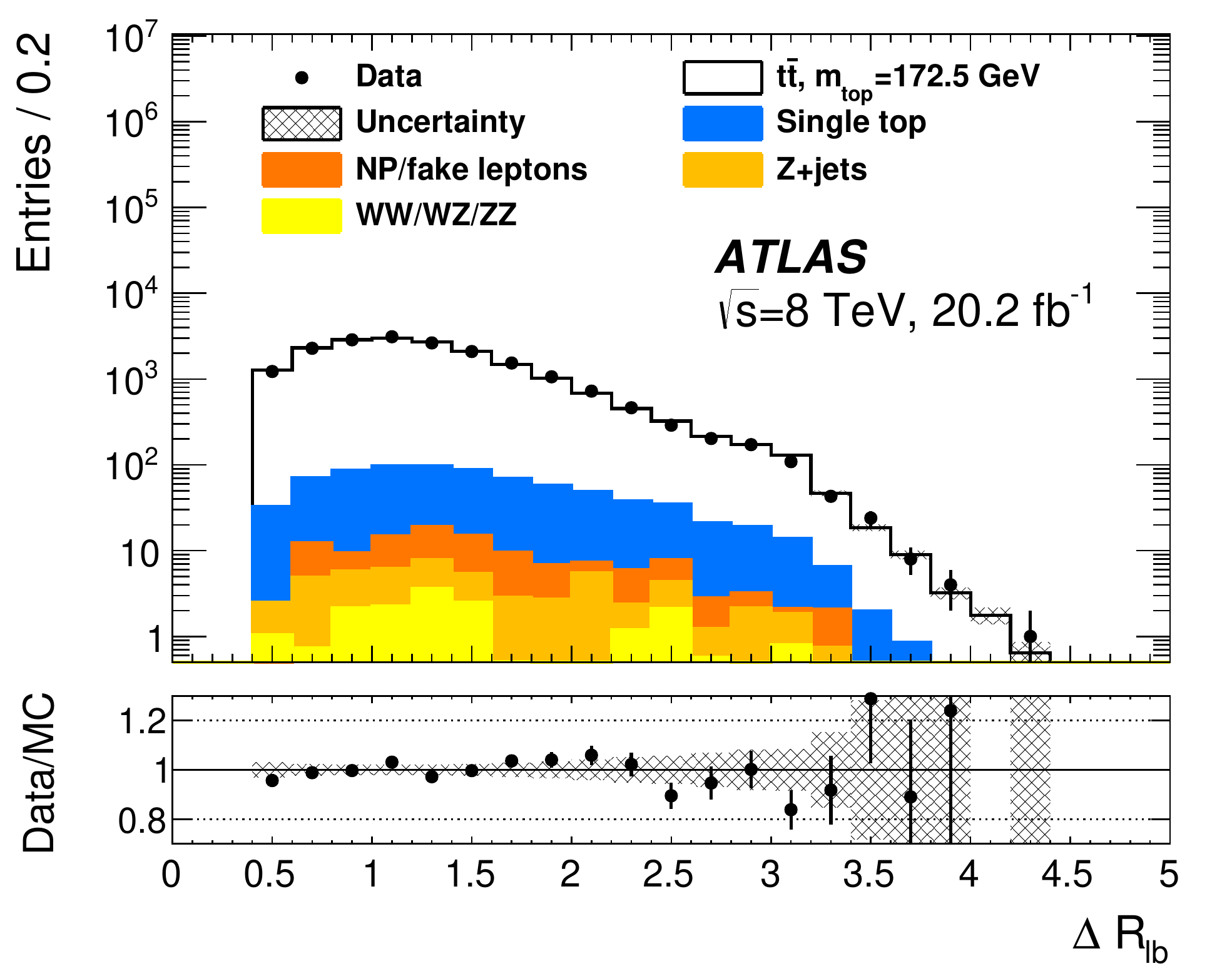}\label{sfig:dataMCdRlb}}
\caption{Kinematic distributions obtained from the objects assigned to the 
 two lepton--\bjet\ pairs for~\protect\subref{sfig:dataMCptlb} the
 pre-selection,
 or~\protect\subref{sfig:dataMCptbj}$-$\protect\subref{sfig:dataMCdRlb} the
 final selection.
 The average \pt\ of the two lepton--\bjet\ pairs, denoted by \ptlb, is shown
 in~\protect\subref{sfig:dataMCptlb}. The \ptlb\ requirement for the final
 selection is indicated by the vertical dashed line.
 The remaining distributions show the \pt\ of the \bjet{s}
 in~\protect\subref{sfig:dataMCptbj}, the \pt\ of the leptons
 in~\protect\subref{sfig:dataMCptle}, and the \dRlb\ of the lepton and
 the \bjet\ for the two lepton--\bjet\ pairs
 in~\protect\subref{sfig:dataMCdRlb}.
 The rightmost bin contains the overflow, if present. For all distributions, the
 number of predicted events is normalised to the one observed in the data.  The
 hatched area corresponds to the statistical uncertainties in the prediction,
 the uncertainty bars to the statistical uncertainties in the data.
 For each figure, the ratio of data and prediction is also presented.
\label{fig:dataMC}}
\end{figure*}
 A matched object is defined as a reconstruction-level object that falls
 within \dR\ of any generator-level object of that type, and a correct match
 means that this generator-level object is the one it originated from.
 Due to acceptance losses and reconstruction inefficiency, not all
 reconstruction-level objects can successfully be matched to their
 generator-level counterparts, resulting in unmatched events.
 The matching efficiency is the fraction of correctly matched events among all
 the matched events, and the \selPurity\ is the fraction of correctly matched
 events among all events, regardless of whether they could be matched or not.
 The corresponding numbers for $\mt=172.5$~\GeV\ are reported
 in \Tab{\ref{tab:selections}}.

 Starting from this pre-selection, an optimisation of the total uncertainty
 in \mt\ is performed.
 A phase-space restriction based on the average \pt\ of the two lepton--\bjet\
 pairs~(\ptlb) is used to obtain the smallest total uncertainty in \mt.
 The corresponding \ptlb\ distribution is shown in \Fig{\ref{sfig:dataMCptlb}}.
 The smallest uncertainty in \mt\ corresponds to $\ptlb >120$~\GeV.
 The difference in shape between data and prediction is covered by the
 systematic uncertainty as detailed in \Sect{\ref{sec:unc}}.
 This restriction is found to also increase the fraction of correctly matched
 events in the \ttbar\ sample, and reduces the number of unmatched or wrongly
 matched events.

 To perform the template parameterisation described in \Sect{\ref{sec:result}},
 an additional selection criterion is applied, restricting the
 reconstructed \mlb\ value (\mlbr) to the range $30~\GeV < \mlbr < 170~\GeV$.
 Applying both restrictions, the numbers of predicted and observed events
 resulting from the final selection are reported in \Tab{\ref{tab:selections}}.
 Using this optimisation, the matching efficiency and the sample purity are much
 improved as reported in the bottom rows of \Tab{\ref{tab:selections}}, while
 retaining about $\OptiFraction\%$ of the events.
 Using this selection, and the objects assigned to the two lepton--\bjet\ pairs,
 the kinematic distributions in the data are well described by the predictions,
 as shown in \Fig{\ref{fig:dataMC}} for the transverse momenta of \bjet{s} and
 leptons, and for the \dRlb\ of the two lepton--\bjet\ pairs.

\section{Template fit and results in the data}
\label{sec:result}
%
\begin{figure*}[tbp!]
\centering
\subfloat[Signal templates in simulation]{
  \includegraphics[width=0.49\textwidth]{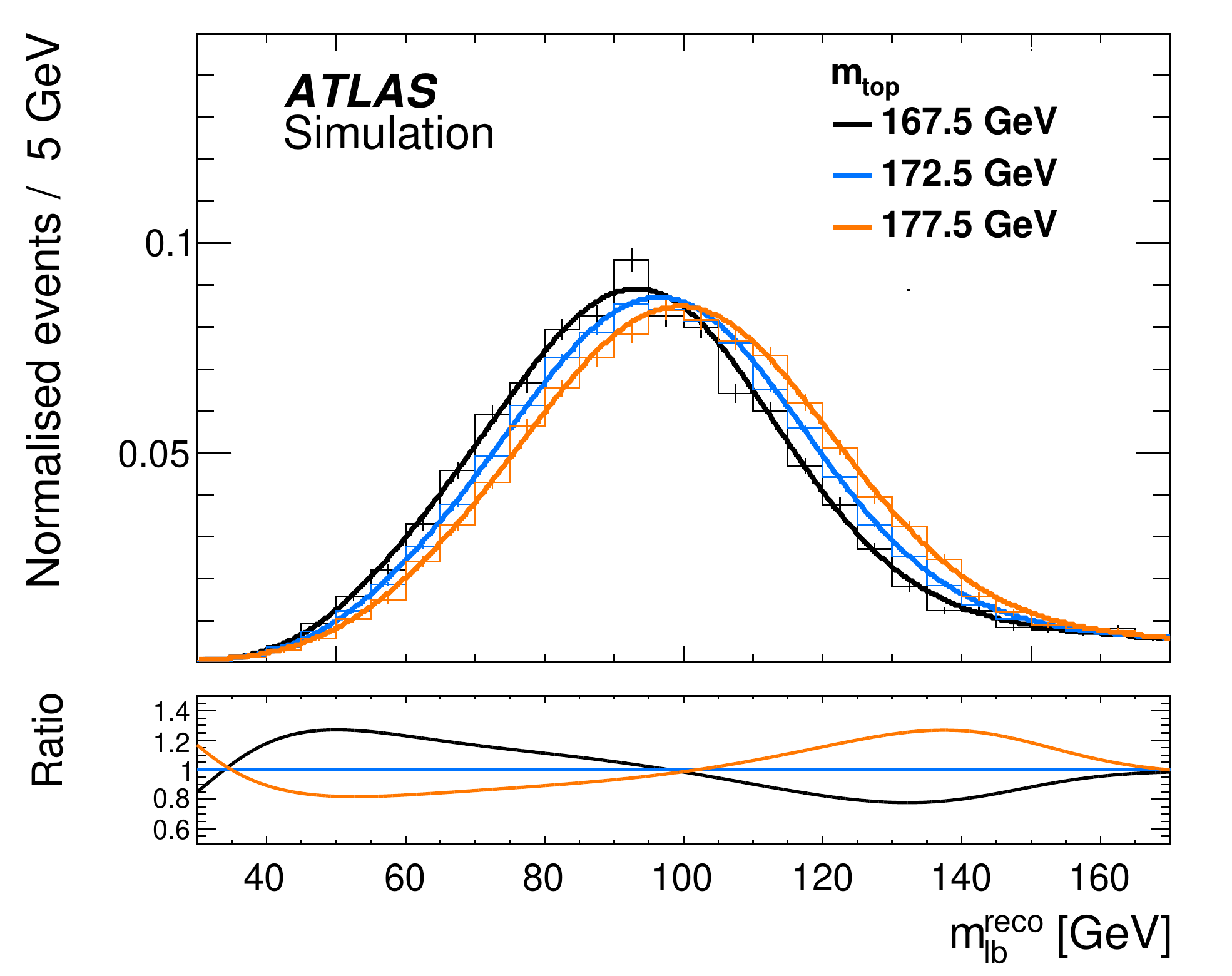}
  \label{sfig:restmpl}
}
\subfloat[\mlbr\ in data and simulation]{
  \includegraphics[width=0.49\textwidth]{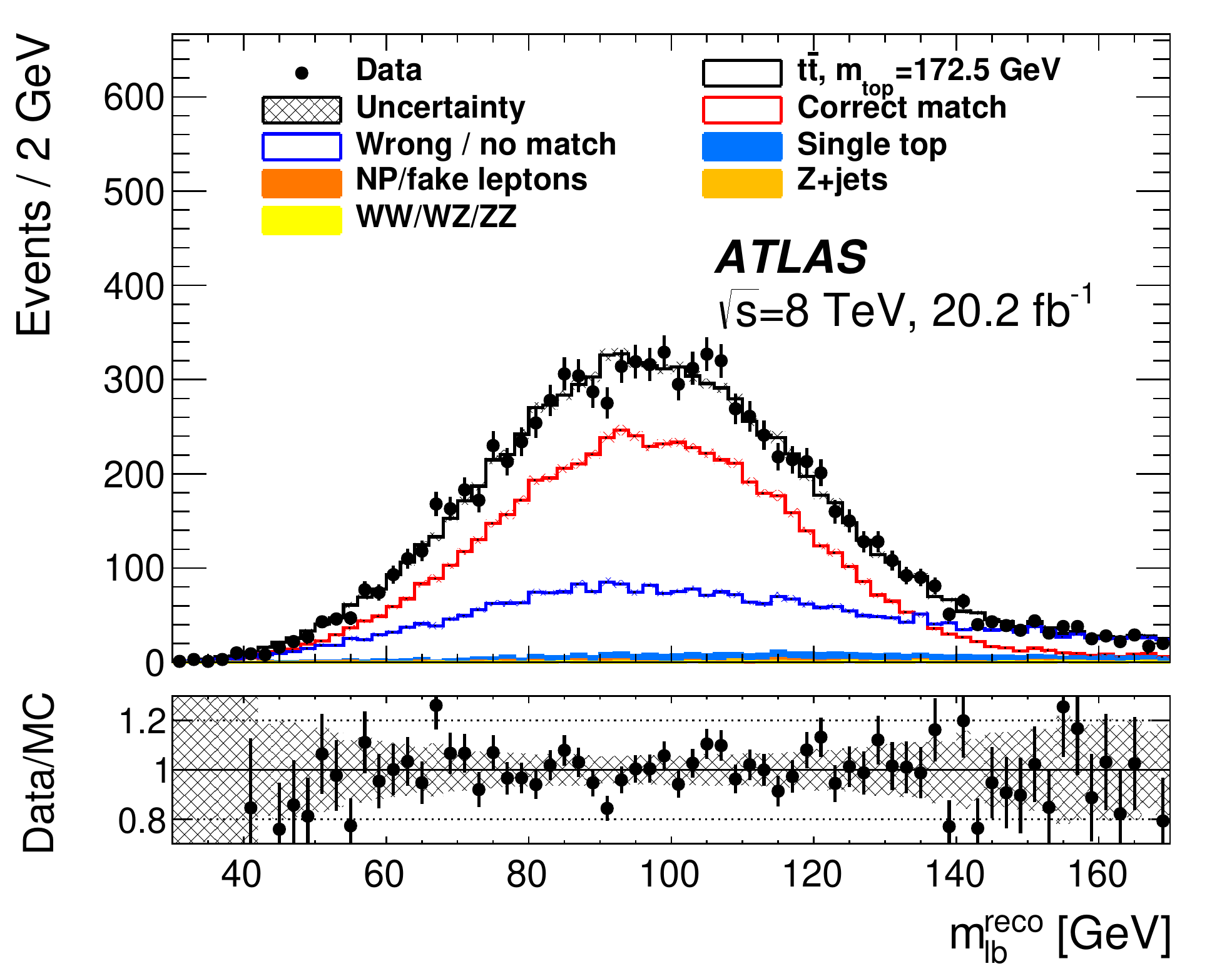}
  \label{sfig:resmlbr}
}
\hfill
\subfloat[Fit to \mlbr\ in data]{
  \includegraphics[width=0.49\textwidth]{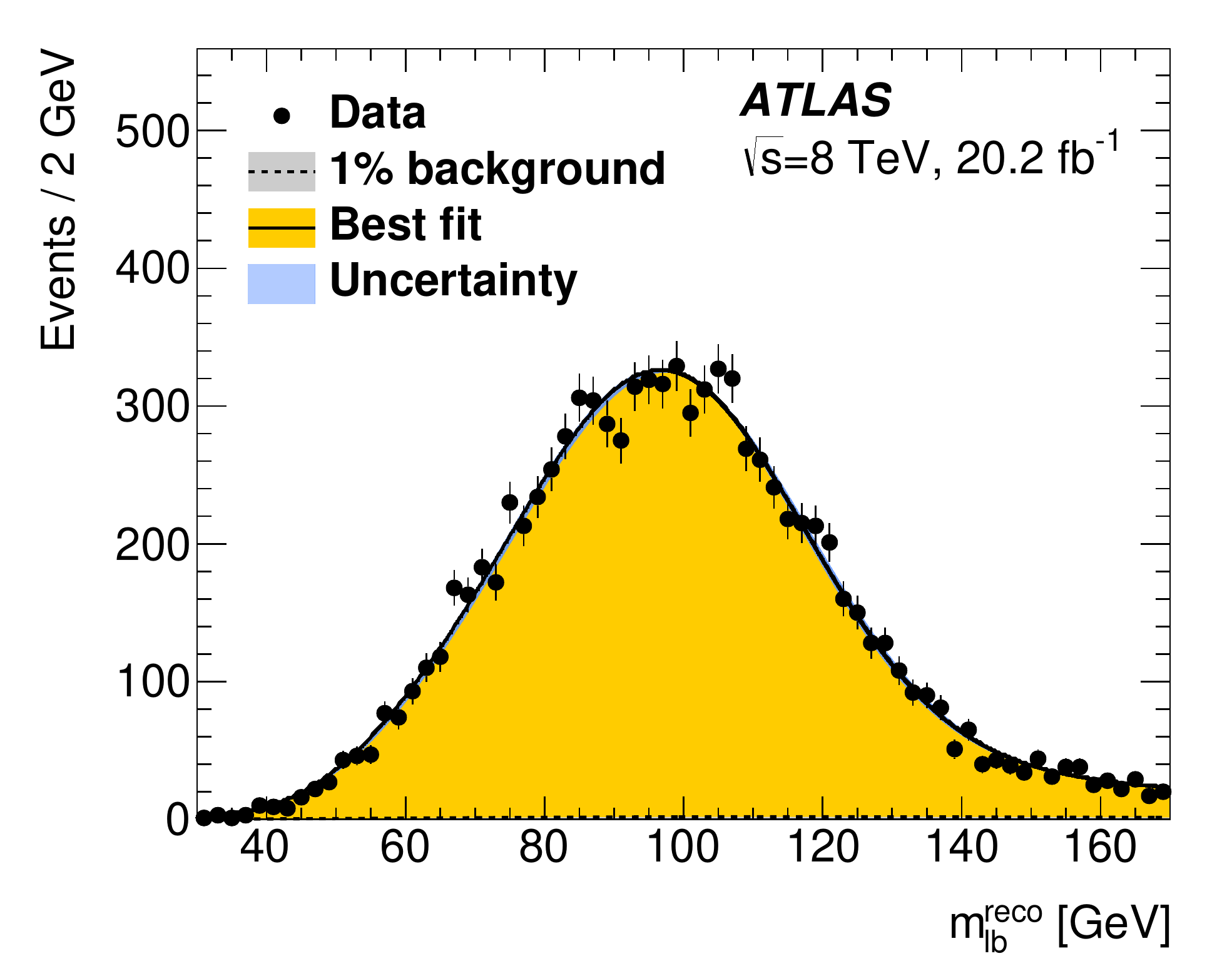}
  \label{sfig:resfits}
}
\subfloat[Logarithm of the likelihood]{
  \includegraphics[width=0.49\textwidth]{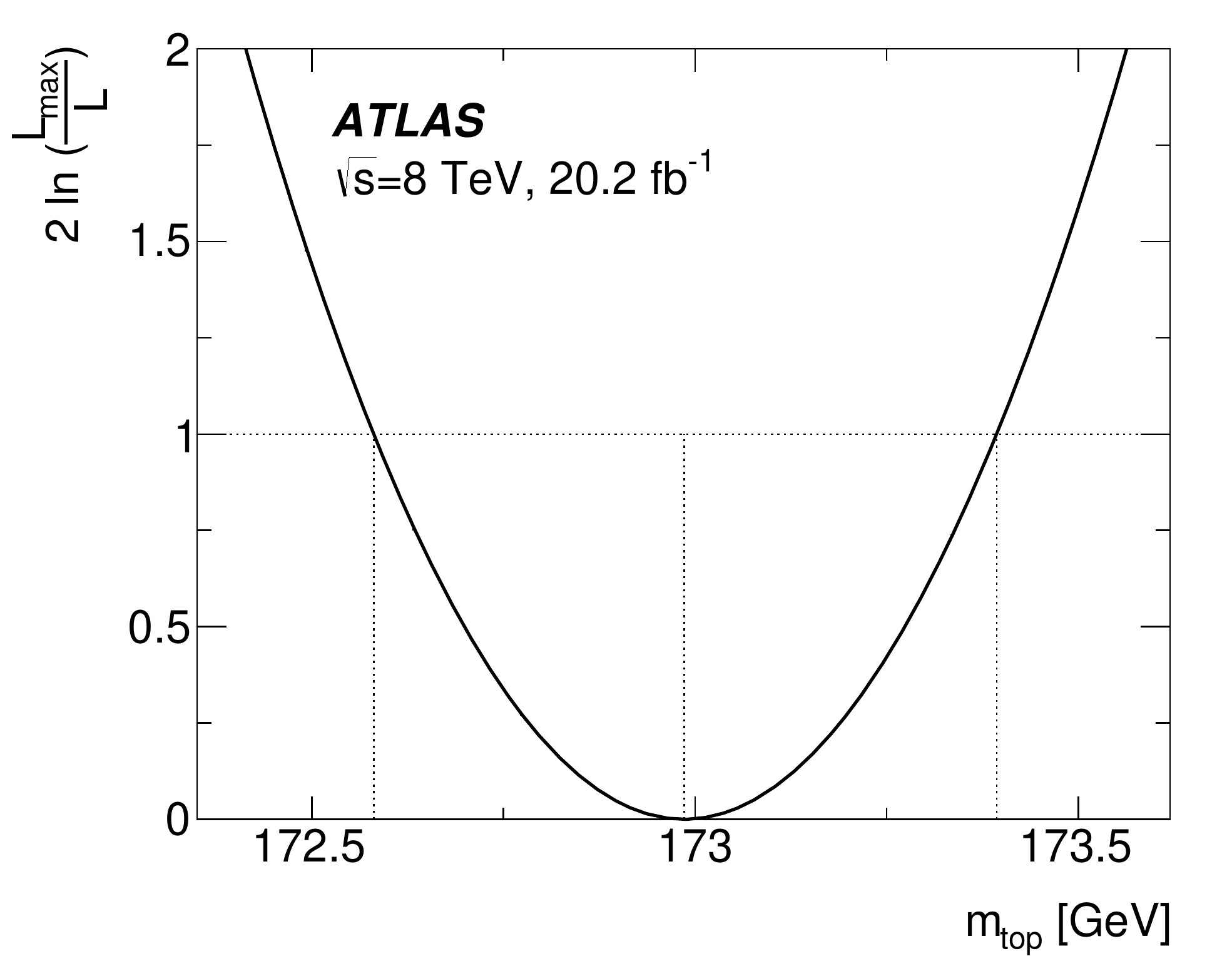}
  \label{sfig:reslike}
}
\caption{Simulated signal templates (histograms) for different values of
  \mt\ together with the template fits (curves) are given
  in~\protect\subref{sfig:restmpl}. The \mlbr\ distribution observed in data in
  comparison to the prediction is shown in~\protect\subref{sfig:resmlbr}. Both
  figures show statistical uncertainties only.
  In~\protect\subref{sfig:resmlbr} the background contributions are too small to
  be distinguished.
  The \mlbr\ distribution is shown in~\protect\subref{sfig:resfits} for data
  with statistical uncertainties together with the fitted probability density
  functions for the background alone (barely visible at the bottom of the
  figure) and for the sum of signal and background. The uncertainty band
  corresponds to the total uncertainty in \mt.
  Finally, the corresponding logarithm of the likelihood as a function of
  \mt\ is displayed in~\protect\subref{sfig:reslike}.
\label{fig:res}}
\end{figure*}
%
 The implementation of the template method used in this analysis is described in
 \Ref{\cite{Aad:2015nba}}.
 For this analysis, the templates are simulated distributions of \mlbr,
 constructed for a number of discrete values of \mt.
 Appropriate functions are fitted to these templates, interpolating between
 different input \mt.
 The remaining parameters of the functions are fixed by a simultaneous fit to
 all templates, imposing linear dependences of the parameters on \mt.
 The resulting template fit function has \mt\ as the only free parameter and an
 unbinned likelihood maximisation gives the value of \mt\ that best describes
 the data.
 Statistically independent signal templates, comprising \ttbar\ and
 single-top-quark events, are constructed as a function of the top quark mass
 used in the MC generator.
 Within the statistical uncertainties, the sum of a Gaussian distribution and a
 Landau function gives a good description of the shape of the
 \mlbr\ distribution as shown in \Fig{\ref{sfig:restmpl}} for three values of
 \mt.
 With this signal choice, the background distribution is independent of \mt, and
 a Landau function is fitted to it.
 The sum of the signal template at $\mt=172.5$~\GeV\ and the background is
 compared to data in \Fig{\ref{sfig:resmlbr}}. It gives a good description of
 the data except for differences that can be accounted for by a different top
 quark mass. In this distribution, the correctly matched events are concentrated
 in the central part, whereas the remainder is less peaked and accounts for most
 of the tails.

 In this analysis the expected statistical precision as well as all systematic
 uncertainties are obtained from pseudo-experiments generated from MC simulated
 samples mimicking ATLAS data.
 To verify the internal consistency of the method, 1000 pseudo-experiments per
 mass point are performed, correcting for oversampling~\cite{BarlowCF}.
 Within uncertainties, and for all \mt\ values, the residuals and pull means are
 consistent with zero and the pull widths are consistent with unity, i.e.~the
 estimator is unbiased and uncertainties are calculated properly.
 The expected statistical uncertainty is obtained from the distribution of the
 statistical uncertainty in the fitted \mt\ of the pseudo-experiments. 
 For $\mt=172.5$~\GeV\ and the data luminosity it amounts to
 $\ExpstatuncEight\pm\ExpstatuncEightunc$~\GeV, where the quoted precision is
 statistical.
 The \mlbr\ distribution in the data is shown in \Fig{\ref{sfig:resfits}}
 together with the corresponding fitted probability density functions for the
 background alone and for the sum of signal and background.
 The value obtained fixing the background contribution to its prediction is $\mt
 =\XZst{\EightValue}{\EightStat}$\,\GeV.
 The statistical uncertainty in \mt\ is taken from the parabolic approximation
 of the logarithm of the likelihood as shown in \Fig{\ref{sfig:reslike}}. The
 observed and predicted values of the statistical uncertainty agree.

\section{Uncertainties affecting the \boldmath$\mt$ determination}
\label{sec:unc}
 The same systematic uncertainty sources as in \Ref{\cite{Aad:2015nba}} are
 investigated.
 Their impact on the analysis is mostly evaluated from pairs of samples
 expressing a particular systematic uncertainty, by constructing the
 corresponding templates and measuring the average difference in \mt\ of the
 pair from 1000 pseudo-experiments.
 To facilitate a combination with other results, every systematic uncertainty is
 assigned a statistical uncertainty, taking into account the statistical
 correlation of the considered samples.
 Following \Ref{\cite{Barlow:2002yb}}, the resulting uncertainty components are
 given in \Tab{\ref{tab:results}} irrespective of their statistical
 significance.
 The uncertainty sources are constructed so as to be uncorrelated with each
 another and thus the total uncertainty squared is calculated as the sum in
 quadrature of all components.
%
\begin{sidewaystable*}[tbp!]
\small
\begin{center}
\begin{tabular}{|l|r|r|r||r|r|r||r|r|r|} \cline{2-10}
\multicolumn{1}{c|}{} & \multicolumn{2}{c|}{$\sqrts=7$~\TeV} & $\sqrts=8$~\TeV & \multicolumn{3}{c||} {Correlations} & 
\multicolumn{3}{c|}{Combinations} \\ \cline{2-10}
\multicolumn{1}{c|}{} & \mtlj\ [\GeV] & \mtdl\ [\GeV]        & \mtdl\ [\GeV]    & $\rho_{01} $ & $\rho_{02}$ & $\rho_{12}$ &
\mtseven\ [\GeV] & \mtdl\ [\GeV] & \mtall\ [\GeV] \\ \hline
Results                               &          172.33 &               173.79 &           \EightValue
                                      &                 &                      &
                                      &          172.99 &         \CombValueDL &            \CombValue \\ \hline
Statistics                            &            0.75 &                 0.54 &            \EightStat
                                      &             $0$ &                  $0$ &                   $0$
                                      &            0.48 &          \CombStatDL &             \CombStat \\
Method                                & 0.11 $\pm$ 0.10 &      0.09 $\pm$ 0.07 &       0.05 $\pm$ 0.07
                                      &             $0$ &                  $0$ &                   $0$
                                      &            0.07 &                 0.05 &                  0.05 \\ \hline
Signal Monte Carlo generator          & 0.22 $\pm$ 0.21 &      0.26 $\pm$ 0.16 &       0.09 $\pm$ 0.15
                                      &         $+1.00$ &              $+1.00$ &               $+1.00$
                                      &            0.24 &                 0.10 &                  0.14 \\
Hadronisation                         & 0.18 $\pm$ 0.12 &      0.53 $\pm$ 0.09 &       0.22 $\pm$ 0.09
                                      &         $+1.00$ &              $+1.00$ &               $+1.00$
                                      &            0.34 &                 0.24 &                  0.23 \\
Initial- and final-state QCD radiation& 0.32 $\pm$ 0.06 &      0.47 $\pm$ 0.05 &       0.23 $\pm$ 0.07
                                      &         $-1.00$ &              $-1.00$ &               $+1.00$
                                      &            0.04 &                 0.24 &                  0.08 \\
Underlying event                      & 0.15 $\pm$ 0.07 &      0.05 $\pm$ 0.05 &       0.10 $\pm$ 0.14
                                      &         $-1.00$ &              $-1.00$ &               $+1.00$
                                      &            0.06 &                 0.10 &                  0.02 \\
Colour reconnection                   & 0.11 $\pm$ 0.07 &      0.14 $\pm$ 0.05 &       0.03 $\pm$ 0.14
                                      &         $-1.00$ &              $-1.00$ &               $+1.00$
                                      &            0.01 &                 0.03 &                  0.01 \\
Parton distribution function          & 0.25 $\pm$ 0.00 &      0.11 $\pm$ 0.00 &       0.05 $\pm$ 0.00
                                      &         $+0.57$ &              $-0.29$ &               $+0.03$
                                      &            0.17 &                 0.04 &                  0.08 \\ \hline
Background normalisation              & 0.10 $\pm$ 0.00 &      0.04 $\pm$ 0.00 &       0.03 $\pm$ 0.00
                                      &         $+1.00$ &              $+0.23$ &               $+0.23$
                                      &            0.07 &                 0.03 &                  0.04 \\
$W/Z$+jets shape                      & 0.29 $\pm$ 0.00 &      0.00 $\pm$ 0.00 &                   $0$
                                      &             $0$ &                      &
                                      &            0.16 &                 0.00 &                  0.09 \\
Fake leptons shape                    & 0.05 $\pm$ 0.00 &      0.01 $\pm$ 0.00 &       0.08 $\pm$ 0.00
                                      &         $+0.23$ &              $+0.20$ &               $-0.08$
                                      &           0.03  &                 0.07 &                  0.05 \\ \hline
Jet energy scale                      & 0.58 $\pm$ 0.11 &      0.75 $\pm$ 0.08 &       0.54 $\pm$ 0.04
                                      &         $-0.23$ &              $+0.06$ &               $+0.35$
                                      &            0.41 &                 0.52 &                  0.41 \\
Relative $b$-to-light-jet energy scale& 0.06 $\pm$ 0.03 &      0.68 $\pm$ 0.02 &       0.30 $\pm$ 0.01
                                      &         $+1.00$ &              $+1.00$ &               $+1.00$
                                      &            0.34 &                 0.32 &                  0.25 \\
Jet energy resolution                 & 0.22 $\pm$ 0.11 &      0.19 $\pm$ 0.04 &       0.09 $\pm$ 0.05
                                      &         $-1.00$ &                  $0$ &                   $0$
                                      &            0.03 &                 0.08 &                  0.08 \\
Jet reconstruction efficiency         & 0.12 $\pm$ 0.00 &      0.07 $\pm$ 0.00 &       0.01 $\pm$ 0.00
                                      &         $+1.00$ &              $+1.00$ &               $+1.00$
                                      &            0.10 &                 0.01 &                  0.04 \\
Jet vertex fraction                   & 0.01 $\pm$ 0.00 &      0.00 $\pm$ 0.00 &       0.02 $\pm$ 0.00
                                      &        $-1.00$  &              $+1.00$ &               $-1.00$
                                      &            0.00 &                 0.02 &                  0.02 \\
\btag                                 & 0.50 $\pm$ 0.00 &      0.07 $\pm$ 0.00 &       0.03 $\pm$ 0.02
                                      &         $-0.77$ &                  $0$ &                   $0$
                                      &            0.25 &                 0.03 &                  0.15 \\
Leptons                               & 0.04 $\pm$ 0.00 &      0.13 $\pm$ 0.00 &       0.14 $\pm$ 0.01
                                      &         $-0.34$ &              $-0.52$ &               $+0.96$
                                      &            0.05 &                 0.14 &                  0.09 \\
\met                                  & 0.15 $\pm$ 0.04 &      0.04 $\pm$ 0.03 &       0.01 $\pm$ 0.01
                                      &         $-0.15$ &              $+0.25$ &               $-0.24$
                                      &            0.08 &                 0.01 &                  0.05 \\
Pile-up                               & 0.02 $\pm$ 0.01 &      0.01 $\pm$ 0.00 &       0.05 $\pm$ 0.01
                                      &             $0$ &                  $0$ &                   $0$
                                      &            0.01 &                 0.05 &                  0.03 \\ \hline
Total systematic uncertainty          & 1.03 $\pm$ 0.31 &      1.31 $\pm$ 0.23 & \EightSyst $\pm$ 0.29
                                      &                 &                      &
                                      &            0.77 &          \CombSystDL &              \CombSyst \\ \hline
Total                                 & 1.27 $\pm$ 0.33 &      1.41 $\pm$ 0.24 &  \EightTot $\pm$ 0.29
                                      &         $-0.07$ & \CorrEightDilSevenLj &              \CorrDil
                                      &            0.91 &           \CombTotDL &              \CombTot \\ \hline
\end{tabular}
\end{center}
\caption{The three measured values of \mt\ together with their statistical
 and systematic uncertainty components are shown on the left.
 The middle part reports the estimated correlations $\rho_\mathrm{ij}$ per pair
 of measurements, with $0$, $1$ and $2$ denoting the \ljets\ and \dil\
 measurements at $\sqrts=7$~\TeV~(from Ref.~\cite{Aad:2015nba}) and the \dil\
 measurement at $\sqrts=8$~\TeV, respectively.
 Finally, the right part lists the \mt\ results for the combinations of the two
 measurements at $\sqrts=7$~\TeV, the two measurements in the \dil\ channel and
 all measurements.
 For the individual measurements, the systematic uncertainty in \mt\ and its
 associated statistical uncertainty is given for each source of uncertainty.
 Assigned correlations are given as integer values, determined correlations as
 real values.
 The last line refers to the sum in quadrature of the statistical and systematic
 uncertainty components or the total correlations, respectively.
\label{tab:results}}
\end{sidewaystable*}
%
 The various sources of systematic uncertainties and the evaluation of their
 effect on \mt\ are briefly described in the following. The values are given
 in \Tab{\ref{tab:results}}.

 {\bf Method:} The mean value of the differences between the fitted and
 generated \mt\ for the MC samples at various input top quark masses is assigned
 as the method calibration uncertainty.
 This also covers effects from limited numbers of MC simulated events in the
 templates.\\
 {\bf Signal Monte Carlo generator:} The difference in \mt\ between the event
 sample produced with the \Mcatnlo\ program~\cite{FRI-0201,FRI-0301} and the
 default \Powheg\ sample, both generated at $\mt=172.5$~\GeV\ and using
 the \Herwig\ program for parton shower, hadronisation and underlying event, is
 quoted as a systematic uncertainty.\\
 {\bf Hadronisation:} The difference in \mt\ between samples produced with
 the \PowhegBox\ program and showered with either the \Pythiasix\ program using
 the P2011C tune or the \Herwig\ and \Jimmy\ programs using the ATLAS AUET2
 tune~\cite{ATL-PHYS-PUB-2011-008} is quoted as a systematic uncertainty.
 This includes different approaches in parton-shower modelling and
 hadronisation, namely the Lund string
 model~\cite{Andersson198331,Andersson:1998tv} and the cluster
 model~\cite{Webber:1983if}.
 The difference in shape between data and prediction observed for the \ptlb\
 distribution shown in \Fig{\ref{sfig:dataMCptlb}} is much reduced when using
 the \PowhegHerwig\ sample and therefore covered by this uncertainty.
 As a check to assess the maximum possible difference in \mt\ caused by the
 mismodelling of the \ptlb\ distribution, the predicted distribution is
 reweighted to the data distribution and the fit is repeated.
 The observed difference in \mt\ from the nominal sample is about \difffit~\GeV,
 well below the statistical uncertainty in the data. Consequently, no additional
 uncertainty is applied.
 Finally, the calibration of the JES and bJES, discussed below, is also
 partially based on a comparison of jet energy responses in event samples
 produced with the \Herwigpp~\cite{Bahr:2008pv} and \Pythiasix\ programs.
 However, it was verified~\cite{ATL-PHYS-PUB-2015-042} that the amount of
 double-counting of JES and hadronisation effects for the \ttbarlj\ channel is
 small.\\
 {\bf Initial- and final-state QCD radiation~(ISR/FSR):} The uncertainty due to
 this effect is evaluated by comparing two dedicated samples generated with
 the \PowhegBox\ and \Pythiasix\ programs that differ in several parameters,
 namely: the QCD scale \LambdaQCD, the transverse momentum scale for space-like
 parton-shower evolution $Q^2_\mathrm{max}$ and the \hdamp\
 parameter~\cite{ATL-PHYS-PUB-2014-005}.
 Half the observed difference between the up variation and the down variation is
 quoted as a systematic uncertainty.
 For comparison, using the signal samples generated at $\mt=172.5$~\GeV, and
 only changing the \hdamp\ parameter but using a much larger range, i.e.~from
 $\infty$ to \mt, the measured \mt\ is lowered by $0.23\pm0.13$~\GeV, where the
 uncertainty is statistical.\\
 {\bf Underlying event (UE):} The difference in UE modelling is assessed by
 comparing \Powheg\ samples based on the same partonic events generated with the
 CT10 PDFs.
 The difference in \mt\ for a sample with the Perugia 2012 tune (P2012) and a
 sample with the P2012 mpiHi tune~\cite{Skands} is assigned as a systematic
 uncertainty.\\
 {\bf Colour reconnection (CR):} This systematic uncertainty is estimated using
 samples with the same partonic events as for the UE uncertainty evaluation, but
 with the P2012 tune and the P2012 loCR tune~\cite{Skands} for PS and
 hadronisation.
 The difference in \mt\ is quoted as a systematic uncertainty.\\
 {\bf Parton distribution function (PDF):} The PDF systematic uncertainty is the
 sum in quadrature of three contributions.
 These are: the sum in quadrature of the differences in \mt\ for the
 26~eigenvector variations of the CTEQ PDF~\cite{Pumplin2002} and two
 differences in \mt\ obtained from reweighting the central CT10 PDF set to the
 MSTW2008 PDF~\cite{MAR-0901} and the NNPDF23 PDF~\cite{Ball:2012cx}.\\
 {\bf Background normalisation:} The normalisations are varied simultaneously
 for the MC-based and the data-driven background estimates according to the
 above mentioned uncertainties.\\
 {\bf Background shapes:} Given the negligible uncertainty in the dilepton
 channel observed in \Ref{\cite{Aad:2015nba}}, no shape uncertainty is evaluated
 for the MC-based background.
 For the data-driven background the shape uncertainty is obtained from the
 estimate of fake-lepton events using the matrix
 method~\cite{ATLAS-CONF-2014-058}.\\
 {\bf Jet energy scale (JES):} Mean jet energies are measured with a relative
 precision of about $1\%$ to $4\%$, typically falling with jet \pt\ and rising
 with jet $\vert\eta\vert$~\cite{ATLAS-CONF-2015-057,ATLAS-CONF-2015-017}.
 The large number of subcomponents of the total JES uncertainty are reduced by a
 matrix diagonalisation of the full JES covariance matrix.
 For each of the resulting 25 significant nuisance
 parameters~\cite{ATLASCollaboration2015b} the corresponding uncertainty in \mt\
 is calculated.
 The total JES-induced uncertainty in \mt\ is obtained by the sum in quadrature
 of the results for the subcomponents.\\
 {\bf Relative $b$-to-light-jet energy scale (bJES):} The bJES is an additional
 uncertainty for the remaining differences between \bjet{s}\ and light jets
 after the global JES is applied and therefore the corresponding uncertainty is
 uncorrelated with the JES uncertainty.
 Jets containing $b$-hadrons are assigned an additional uncertainty of $0.2\%$
 to $1.2\%$, with lowest uncertainties for
 high-\pt\ \bjet{s}~\cite{ATLASCollaboration2015b}.\\
 {\bf Jet energy resolution (JER):} The JER uncertainty is determined by the sum
 in quadrature of the \mt\ differences between the varied samples and the
 nominal sample or, where applicable, half the fitted difference between the up
 variation and the down variation of the components of the eigenvector
 decomposition.\\
 {\bf Jet reconstruction efficiency (JRE):} The JRE uncertainty is evaluated by
 randomly removing $2\%$ of the jets with $\pt< 30$~\GeV\ from the MC simulated
 events prior to the event selection to reflect the precision with which the
 data-to-MC JRE ratio is known~\cite{Aad:1409965}.
 The \mt\ difference with respect to the nominal sample is taken as a systematic
 uncertainty.\\
 {\bf Jet vertex fraction (JVF):} When summing the scalar \pt\ of all tracks in
 a jet, the JVF is the fraction contributed by tracks originating at the primary
 vertex. The uncertainty is evaluated by varying the requirement on the JVF
 within its uncertainty~\cite{ATLAS-CONF-2013-083}.\\
 {\bf \boldmath$\btag$:} Mismodelling of the \btag\ efficiency and mistag rate
 is accounted for by the application of scale factors which depend on jet \pt\
 and jet $\eta$ to MC simulated events~\cite{Aad:2015ydr}.
 The eigenvector decomposition~\cite{Aad:2015ydr,ATLAS-CONF-2014-004} accounts
 for the uncertainties in the \btag, $c/\tau$-tagging and mistagging scale
 factors.
 The final \btag\ uncertainty is the sum in quadrature of these uncorrelated
 components.\\
 {\bf Lepton uncertainties:} The lepton uncertainties measured in
 $\jpsi\to \ell\ell$ and $Z\to \ell\ell$ events are related to the electron
 energy or muon momentum scales and resolutions, and the trigger and
 identification
 efficiencies~\cite{CERN-PH-EP-2014-040,CERN-PH-EP-2014-151,ATLASCollaboration2014b}.
 For each component, the corresponding uncertainty is propagated to the analysis
 including the recalculation of the \met.\\
 {\bf Missing transverse momentum (\boldmath$\met$):} The remaining contribution
 to the \met\ uncertainty stems from the uncertainties in calorimeter cell
 energies associated with low-\pt\ jets ($7~\GeV< \pt < 20~\GeV$), without any
 corresponding reconstructed physics object or from pile-up interactions. Their
 impact is accounted for as described in \Ref{\cite{ATLAS-MET-NEW}}.\\
 {\bf Pile-up:} Besides the component treated in the JES, the residual
 dependence of the fitted \mt\ on the amount of pile-up activity and a possible
 MC mismodelling is determined.
 The \mt\ dependence as functions of \nvtx\ and \meanmu\ is found to be
 consistent in data and simulation. The corresponding uncertainty evaluated from
 the remaining difference is small.

 The systematic uncertainties quoted in \Tab{\ref{tab:results}} carry
 statistical uncertainties.
 The statistical precision of a single sample fit is about $100$~\MeV.
 The statistical correlation of the samples is calculated from the fraction of
 shared events.
 Pairs of samples with only a change in a single parameter have high correlation
 and correspondingly low statistical uncertainty in the difference in \mt, while
 a pair of statistically independent samples results in a larger uncertainty.

 In summary, the result in the \dil\ channel at $\sqrts=8$~\TeV\ of
 $\mt=\XZ{\EightValue}{\EightStat}{\EightSyst}$~\GeV\ is
 about \EightImprovementSeven\ more precise than the one obtained from the
 $\sqrts=7$~\TeV\ data and the most precise single result in this decay channel
 to date.
 The increased precision is partly driven by a better knowledge of the JES and
 bJES.
 In addition, the applied optimisation procedure significantly reduces the total
 systematic uncertainty, mostly due to a lower impact of the JES and theory
 modelling uncertainties.

\section{Combination with previous ATLAS measurements}
\label{sec:comb}
 The combination of the \mt\ results follows the approach developed for the
 combination of the $\sqrts=7$~\TeV\ measurements in \Ref{\cite{Aad:2015nba}}
 including the evaluation of the correlations.
 For combining the measurements from data at different \cmes\ a mapping of
 uncertainty categories is performed.
 Complex cases are the uncertainty components involving eigenvector
 decompositions such as the JES, the JER and the \btag\ scale factor
 uncertainties.
 The $\sqrts=7$ and $8$~\TeV\ measurements are treated as uncorrelated for the
 nuisance parameters of the JER and the \btag, $c/\tau$-tagging and mistagging
 uncertainties. A correlated treatment of the estimators for the flavour-tagging
 nuisance parameters results in an insignificant change in the combination.
 The total JES uncertainty consists of about 20 eigenvector components, which
 partly differ for the analyses of $\sqrts=7$ and $8$~\TeV\ data, which make use
 of the EM+JES and the LCW+GSC~\cite{ATLAS-CONF-2015-057} jet calibrations,
 respectively.
 For the combination, a mapping between uncertainty components at the
 different \cmes\ is employed to identify the corresponding ones. The
 combination was found to be stable against variations of the assumptions for
 ambiguous cases.

 The combination is performed using the best linear unbiased estimate~(BLUE)
 method~\cite{BLUE1,BLUERN}, implying Gaussian probability density functions for
 all uncertainties, using the implementation described in \Ref{\cite{BLUEcpp}}.
 The central values, the list of uncertainty components and the correlations
 $\rho$ of the estimators for each uncertainty component have to be provided.
 For the statistical, method calibration, MC-based background shape at
 $\sqrts=7$~\TeV, and pile-up uncertainties in \mt\ the measurements are assumed
 to be uncorrelated.
 For the remaining uncertainties in \mt, when using $\pm 1\sigma$ variations of
 a systematic effect, e.g.~when changing the bJES by $\pm 1\sigma$, there are
 two possibilities.
 When simultaneously applying a variation for a systematic uncertainty,
 e.g.~$+1\sigma$ for the bJES to a pair of analyses, e.g.~the \dil\ measurements
 at $\sqrts=7$ and $8$~\TeV, both analyses can result in a larger or
 smaller \mt\ value than what is obtained for the nominal case (full
 correlation, $\rho=+1$), or one analysis can obtain a larger and the other a
 smaller value (full anti-correlation, $\rho=-1$).
 Consequently, an uncertainty from a source only consisting of a single
 variation, such as the uncertainty related to the choice of MC generator for
 signal events, results in a correlation of $\rho=\pm1$.
 The estimator correlations for composite uncertainties are evaluated by adding
 the covariance terms of the subcomponents $i$ with $\rho_i=\pm1$ and dividing
 by the total uncertainties for that source.
 The resulting estimator correlation per uncertainty is quoted
 in \Tab{\ref{tab:results}} and used in the combination.
%
\begin{figure*}[tbp!]
\centering
\subfloat[$\sqrts=7$~\TeV\ versus $\sqrts=8$~\TeV\ \dil]{
  \includegraphics[width=0.49\textwidth]{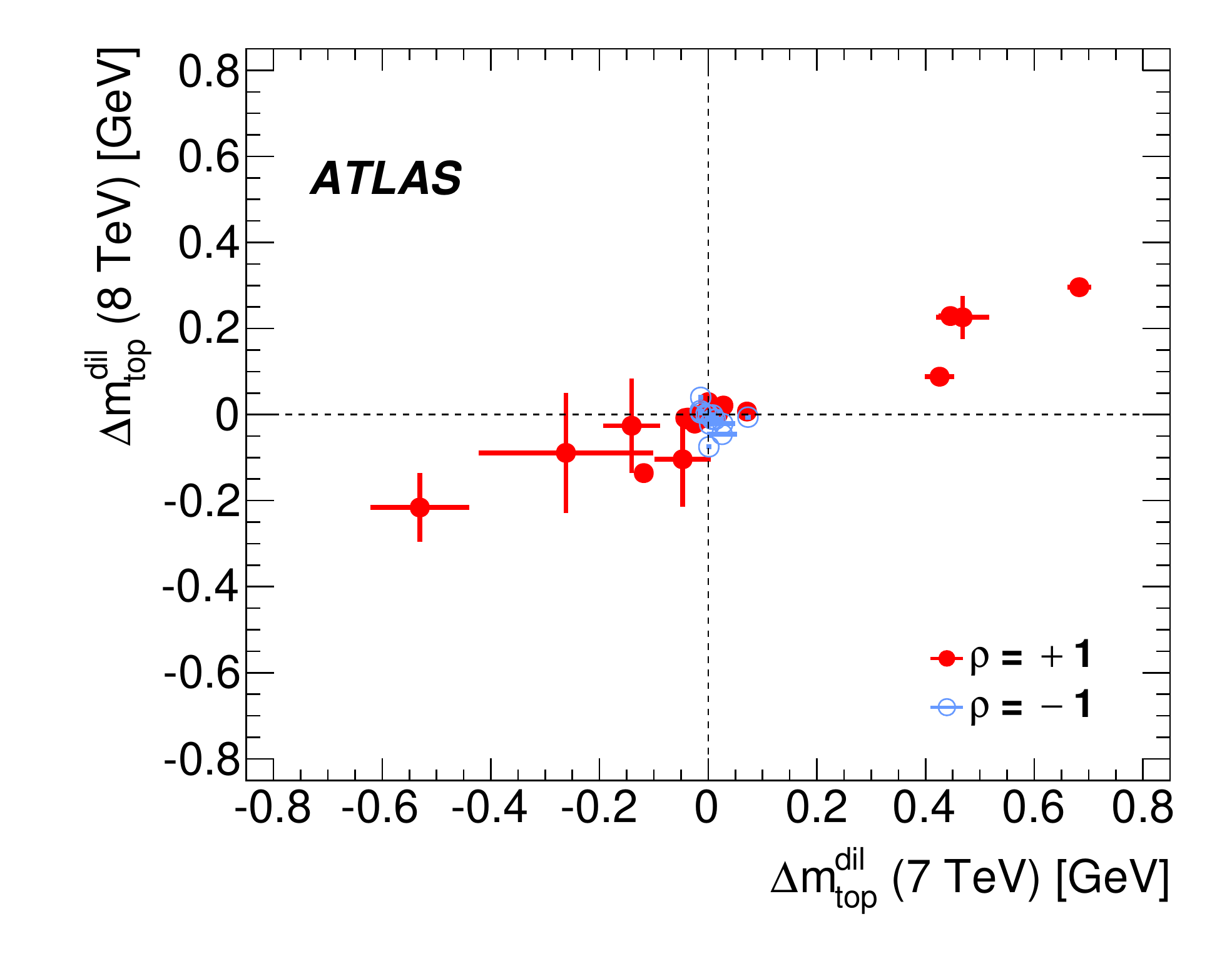}\label{sfig:comr12}}
\subfloat[$\sqrts=7$~\TeV\ \ljets\ versus $\sqrts=8$~\TeV\ \dil]{
  \includegraphics[width=0.49\textwidth]{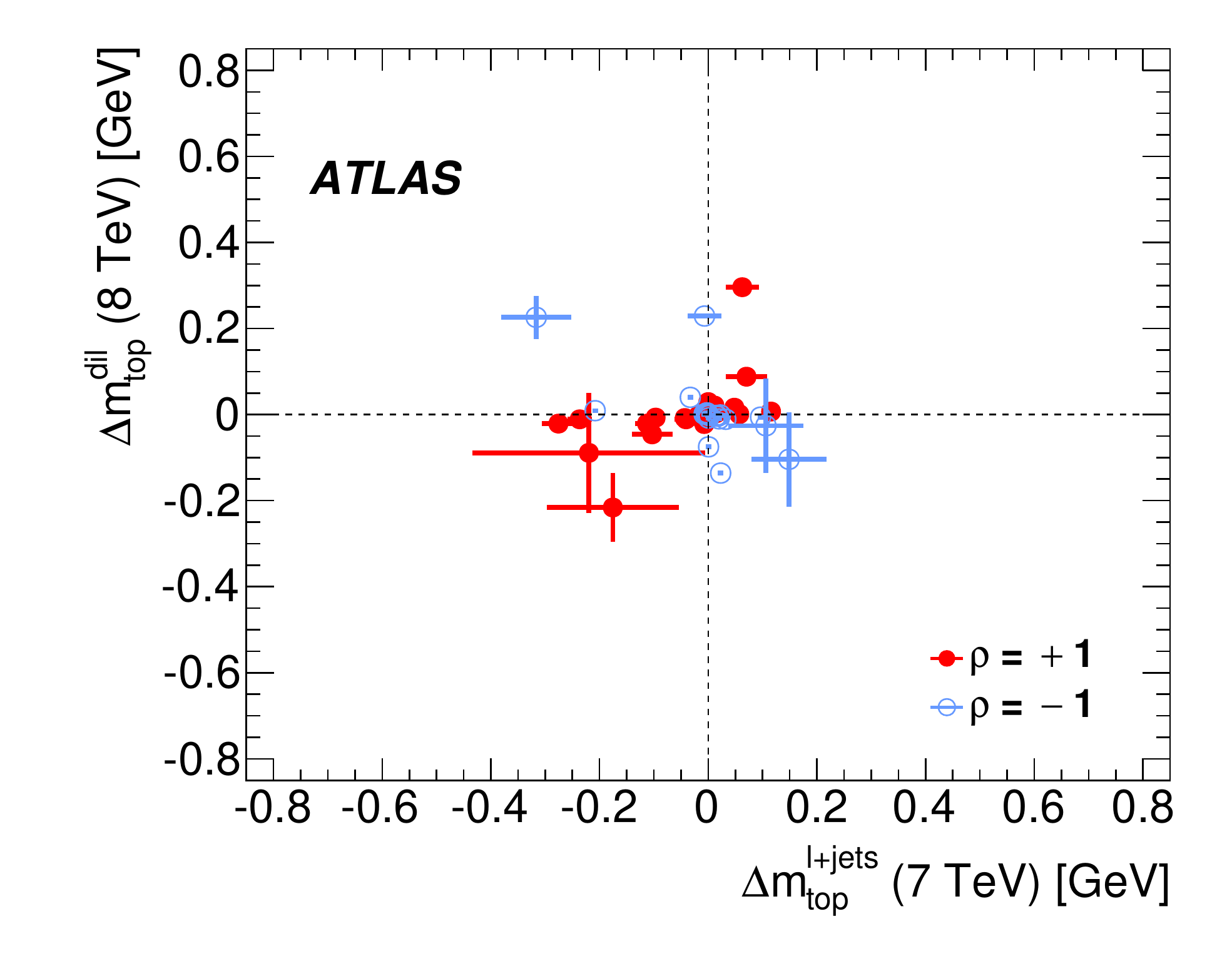}\label{sfig:comr02}}
\caption{The pairwise differences in \mt\ when simultaneously varying both 
 analyses for a systematic uncertainty.
 Each cross indicates the statistical precisions of the systematic uncertainty.
 The red full points indicate $\rho=1$, the blue open points $\rho=-1$.
\label{fig:com}}
\end{figure*}

 The evaluated uncertainties in \mt\ for the uncertainty components for the two
 dilepton analyses, denoted by \dmtdl, are shown in \Fig{\ref{sfig:comr12}}.
 Each point represents a systematic uncertainty together with a cross,
 indicating the respective statistical precision of the systematic uncertainty
 in the two analyses.
 The red full points indicate $\rho=1$, the blue open points $\rho=-1$.
 Given the similarity of the analyses, a positive estimator correlation is
 observed for most uncertainty components of the two measurements in the \dil\
 channel.
 The corresponding distribution for the \ljets\ measurement at $\sqrts=7$~\TeV\
 and the \dil\ measurement at $\sqrts=8$~\TeV\ is given
 in \Fig{\ref{sfig:comr02}}.
 In this figure, the estimates are anti-correlated for several significant
 uncertainties.
 This is caused by the in-situ measurement of the jet energy
 scale factor~(JSF) and relative $b$-to-light-jet energy scale factor~(bJSF) in
 the \threed\ \ljets\ analysis, detailed in \Ref{\cite{Aad:2015nba}}.
 The resulting total correlation for this pair is very low as shown
 in \Tab{\ref{tab:results}}. The combination strongly profits from this.

 The central values of the three measurements, their uncertainty components, the
 determined correlations per pair of measurements and the results of the
 combinations are given in \Tab{\ref{tab:results}}.
 The pairwise differences in the three measurements are $\CompatibSeven\sigma$
 for the $\sqrts=7$~\TeV\ measurements, $\CompatibMixed\sigma$ for the \ljets\
 measurement at $\sqrts=7$~\TeV\ and the \dil\ measurement at $\sqrts=8$~\TeV\
 and $\CompatibDil\sigma$ for the two \dil\ measurements. For all three cases
 $\sigma$ denotes the one standard deviation of the respective \mt\ difference.
 The combined result in the \dil\ channel alone is
 $\mtdl=\XZ{\CombValueDL}{\CombStatDL}{\CombSystDL}\GeV
 = \XZtot{\CombValueDL}{\CombTotDL}$~\GeV, providing no significant improvement
 with respect to the more precise result at $\sqrts=8$~\TeV\ which carries a
 BLUE combination weight of \CombWeightDLeightDL. This is a mere consequence of
 the measurement correlation of \CorrDil, which is close to the ratio of
 uncertainties~(see~\Ref{\cite{BLUERN}}).
 The $\chi^2$ probability of the combination is \CombChiSqProbDL.
 The stability of the combination is assessed from the results of 1000
 combinations for which all input uncertainties are varied within their
 statistical uncertainties, which for some cases also result in different
 correlations~(see~\Fig{\ref{fig:com}}).
 The corresponding distributions of the central values and uncertainties of the
 combinations are approximately Gaussian, with a width of \CombValueStabDL~\GeV\
 and of \CombTotStabDL~\GeV, respectively.

 The combination of all three measurements provides a \CombImprovement\
 improvement with respect to the most precise single input measurement.
 The combined result is $\mtall = \XZ{\CombValue}{\CombStat}{\CombSyst}\GeV
 = \XZtot{\CombValue}{\CombTot}$\,\GeV.
 The $\chi^2$ probability of the combination is \CombChiSqProb\ and the BLUE
 combination weights of the \ljets\ and \dil\ measurements at $\sqrts=7$~\TeV\
 and the \dil\ measurement at $\sqrts=8$~\TeV\
 are \CombWeightLLseven, \CombWeightDLseven\ and \CombWeightDLeight,
 respectively.
 Again, the central value and the combined total uncertainty are both stable at
 the level of \CombValueStab~\GeV.

\FloatBarrier
\section{Conclusion}
\label{sec:conclusion}
 The top quark mass is measured in the \ttbarll\ channel from about
 $\atlumo$~\invfb\ of $\sqrts=8$~\TeV\ proton--proton collision data recorded by
 the ATLAS detector at the LHC.
 Compared to the latest ATLAS measurement in this decay channel, the event
 selection is refined exploiting the average \pt\ of the lepton--\bjet\ pairs to
 enhance the fraction of correctly reconstructed events, thereby reducing the
 systematic uncertainties.
 Using the optimal point in terms of total uncertainty observed in a phase-space
 scan of this variable as an additional event selection criterion, the measured
 value of \mt\ is
%
\begin{eqnarray*}
  \mt &=&\XZ{\EightValue}{\EightStat}{\EightSyst}\GeV,
\end{eqnarray*}
%
 with a total uncertainty of \EightTot~\GeV.
 The precision is mainly limited by systematic uncertainties, mostly by the
 calibration of the jet energy scale, and to a lesser extent by the calibration
 of the relative $b$-to-light-jet energy scale and by the Monte Carlo modelling
 of signal events.

 This measurement is combined with the ATLAS measurements in the \ttbarlj\
 and \ttbarll\ decay channels from $\sqrts = 7$~\TeV\ data.
 The correlations of the measurements are evaluated for all sources of the
 systematic uncertainty.
 Using a dedicated mapping of uncertainty categories, the combination of the
 three measurements results in
%
\begin{eqnarray*}
 \mt &=& \XZ{\CombValue}{\CombStat}{\CombSyst}\GeV,
\end{eqnarray*}
%
 with a total uncertainty of \CombTot~\GeV, i.e.~a relative precision of
 $0.4\%$.
 The result is mostly limited by the calibration of the jet energy scales and by
 the Monte Carlo modelling of signal events.

\section*{Acknowledgements}

We thank CERN for the very successful operation of the LHC, as well as the
support staff from our institutions without whom ATLAS could not be
operated efficiently.

We acknowledge the support of ANPCyT, Argentina; YerPhI, Armenia; ARC, Australia; BMWFW and FWF, Austria; ANAS, Azerbaijan; SSTC, Belarus; CNPq and FAPESP, Brazil; NSERC, NRC and CFI, Canada; CERN; CONICYT, Chile; CAS, MOST and NSFC, China; COLCIENCIAS, Colombia; MSMT CR, MPO CR and VSC CR, Czech Republic; DNRF and DNSRC, Denmark; IN2P3-CNRS, CEA-DSM/IRFU, France; GNSF, Georgia; BMBF, HGF, and MPG, Germany; GSRT, Greece; RGC, Hong Kong SAR, China; ISF, I-CORE and Benoziyo Center, Israel; INFN, Italy; MEXT and JSPS, Japan; CNRST, Morocco; FOM and NWO, Netherlands; RCN, Norway; MNiSW and NCN, Poland; FCT, Portugal; MNE/IFA, Romania; MES of Russia and NRC KI, Russian Federation; JINR; MESTD, Serbia; MSSR, Slovakia; ARRS and MIZ\v{S}, Slovenia; DST/NRF, South Africa; MINECO, Spain; SRC and Wallenberg Foundation, Sweden; SERI, SNSF and Cantons of Bern and Geneva, Switzerland; MOST, Taiwan; TAEK, Turkey; STFC, United Kingdom; DOE and NSF, United States of America. In addition, individual groups and members have received support from BCKDF, the Canada Council, CANARIE, CRC, Compute Canada, FQRNT, and the Ontario Innovation Trust, Canada; EPLANET, ERC, FP7, Horizon 2020 and Marie Sk{\l}odowska-Curie Actions, European Union; Investissements d'Avenir Labex and Idex, ANR, R{\'e}gion Auvergne and Fondation Partager le Savoir, France; DFG and AvH Foundation, Germany; Herakleitos, Thales and Aristeia programmes co-financed by EU-ESF and the Greek NSRF; BSF, GIF and Minerva, Israel; BRF, Norway; Generalitat de Catalunya, Generalitat Valenciana, Spain; the Royal Society and Leverhulme Trust, United Kingdom.

The crucial computing support from all WLCG partners is acknowledged
gratefully, in particular from CERN and the ATLAS Tier-1 facilities at
TRIUMF (Canada), NDGF (Denmark, Norway, Sweden), CC-IN2P3 (France),
KIT/GridKA (Germany), INFN-CNAF (Italy), NL-T1 (Netherlands), PIC (Spain),
ASGC (Taiwan), RAL (UK) and BNL (USA) and in the Tier-2 facilities
worldwide.

\bibliographystyle{atlasBibStyleWithTitle}
\bibliography{TOPQ_2016_03,ATLAS}
\newpage
\begin{flushleft}
{\Large The ATLAS Collaboration}

\bigskip

M.~Aaboud$^\textrm{\scriptsize 135d}$,
G.~Aad$^\textrm{\scriptsize 86}$,
B.~Abbott$^\textrm{\scriptsize 113}$,
J.~Abdallah$^\textrm{\scriptsize 64}$,
O.~Abdinov$^\textrm{\scriptsize 12}$,
B.~Abeloos$^\textrm{\scriptsize 117}$,
R.~Aben$^\textrm{\scriptsize 107}$,
O.S.~AbouZeid$^\textrm{\scriptsize 137}$,
N.L.~Abraham$^\textrm{\scriptsize 149}$,
H.~Abramowicz$^\textrm{\scriptsize 153}$
% ...
\bigskip
\\
% ...
$^{12}$ Institute of Physics, Azerbaijan Academy of Sciences, Baku, Azerbaijan\\
% ...
$^{64}$ University of Iowa, Iowa City IA, United States of America\\
% ...
$^{86}$ CPPM, Aix-Marseille Universit{\'e} and CNRS/IN2P3, Marseille, France\\
% ...
$^{107}$ Nikhef National Institute for Subatomic Physics and University of Amsterdam, Amsterdam, Netherlands\\
% ...
$^{113}$ Homer L. Dodge Department of Physics and Astronomy, University of Oklahoma, Norman OK, United States of America\\
% ...
$^{117}$ LAL, Univ. Paris-Sud, CNRS/IN2P3, Universit{\'e} Paris-Saclay, Orsay, France\\
% ...
$^{135}$ $^{(a)}$ Facult{\'e} des Sciences Ain Chock, R{\'e}seau Universitaire de Physique des Hautes Energies - Universit{\'e} Hassan II, Casablanca; $^{(b)}$ Centre National de l'Energie des Sciences Techniques Nucleaires, Rabat; $^{(c)}$ Facult{\'e} des Sciences Semlalia, Universit{\'e} Cadi Ayyad, LPHEA-Marrakech; $^{(d)}$ Facult{\'e} des Sciences, Universit{\'e} Mohamed Premier and LPTPM, Oujda; $^{(e)}$ Facult{\'e} des sciences, Universit{\'e} Mohammed V, Rabat, Morocco\\
% ...
$^{137}$ Santa Cruz Institute for Particle Physics, University of California Santa Cruz, Santa Cruz CA, United States of America\\
% ...
$^{149}$ Department of Physics and Astronomy, University of Sussex, Brighton, United Kingdom\\
% ...
$^{153}$ Raymond and Beverly Sackler School of Physics and Astronomy, Tel Aviv University, Tel Aviv, Israel\\
% ...
$^{a}$ Also at Department of Physics, King's College London, London, United Kingdom\\
$^{b}$ Also at Institute of Physics, Azerbaijan Academy of Sciences, Baku, Azerbaijan\\
$^{c}$ Also at Novosibirsk State University, Novosibirsk, Russia\\
$^{d}$ Also at TRIUMF, Vancouver BC, Canada\\
$^{e}$ Also at Department of Physics {\&} Astronomy, University of Louisville, Louisville, KY, United States of America\\
% ...
\end{flushleft}


\end{document}